%% file: neurips_2026.tex
\documentclass{article}

\usepackage[sglblindworkshop, final]{neurips_2026}
\workshoptitle{PTA: From Pretrained Representations to Acting Agents}

\usepackage[utf8]{inputenc}
\usepackage[T1]{fontenc}
\usepackage{hyperref}
\usepackage{url}
\usepackage{booktabs}
\usepackage{amsfonts}
\usepackage{amsmath}
\usepackage{amssymb}
\usepackage{amsthm}
\usepackage{nicefrac}
\usepackage{microtype}
\usepackage{xcolor}
\usepackage{graphicx}
\usepackage{multirow}
\usepackage{array}
\usepackage{algorithm}
\usepackage{algorithmic}
\usepackage{tabularx}
\usepackage{array}
\usepackage{makecell}

\title{
Collective Intelligence or Collective Error? Social-Epistemic Dynamics in Diverse Multi-Agent Societies
}

\author{
  Akash Das \\
  \texttt{akash.das@fmr.com}
  \And
  Ishan Roy \\
  \texttt{ishan.roy@fmr.com}
}

\begin{document}

\maketitle

\begin{abstract}
Multi-agent systems increasingly rely on societies of language-model agents
that differ in capability, reasoning behavior, functional role, and social
position. While such diversity can provide complementary information,
interaction can also cause agents to overweight socially influential
information and amplify collective error. We study this problem through a
social-epistemic framework that treats collective reasoning as an evolving
process shaped jointly by agent diversity, interaction topology, social
status, and the alignment between social influence and epistemic competence.
Across controlled experiments spanning heterogeneous language models,
multiple reasoning benchmarks, and the interactive ALFWorld environment,
we systematically vary these factors while separating evidence-driven
belief updates from source-driven social influence. Our results show that diversity alone does not reliably produce collective
intelligence: unstructured heterogeneity yields only modest and inconsistent
gains, while competence-aligned social influence produces substantially
larger improvements in collective accuracy. In contrast, when social
influence is concentrated on less competent agents, repeated interaction
systematically degrades performance and can erase useful epistemic
disagreement. The effect persists in sequential decision making, where
competence-aligned interaction improves task success while adversarial
influence amplifies errors. Mechanistically, we find that source status
changes the weighting of otherwise identical evidence, with high-status
sources receiving disproportionately greater weight when their evidence is
weak or incorrect. This source-dependent distortion compounds over repeated
interaction and provides a concrete pathway from local social bias to
collective error. Finally, an evidence-gated updating mechanism selectively
reduces socially induced updating when available evidence conflicts with
source influence, substantially recovering performance from the adversarial
regime without eliminating interaction altogether. Together, these results
show that the reliability of a multi-agent society depends not simply on
how capable or diverse its agents are, but on whether its social structure
amplifies epistemically reliable information.
\end{abstract}

\section{Introduction}
\label{sec:introduction}

Large language models are increasingly being used not as isolated systems, but as teams of interacting agents. Instead of relying on a single model to solve a problem from beginning to end, multi-agent systems allow several agents to generate solutions, exchange information, challenge one another, and reach a shared decision. This approach is attractive because different agents may bring different capabilities and different views of the same problem. When these differences are useful, interaction can allow a group to solve problems that individual agents may struggle with on their own. A central idea behind this promise is \textit{diversity}. Agents can differ in their underlying models, their ways of reasoning, or the roles they play within a system. Such differences can create disagreement, but disagreement is not necessarily undesirable. If different agents make different errors, one agent may provide information that helps another correct its mistake. A society in which agents maintain different hypotheses can therefore contain more useful information than a society in which every agent immediately produces the same answer.

The situation changes once these agents begin interacting. An agent does not only receive the content of another agent's answer; it also receives signals about the source of that answer. A highly confident agent may appear more convincing than an uncertain one. An agent that has performed well in previous tasks may be trusted more strongly. A designated expert or senior agent may receive greater attention than a peer. Similarly, an agent that occupies a central position in an interaction network may have more opportunities to influence the rest of the group. These signals can affect how information is weighted even when the underlying evidence remains unchanged, creating a problem that is not captured by accuracy alone. Consider a society in which several agents independently produce different answers, one of which is correct. If interaction allows the agents to compare their evidence and identify the correct answer, disagreement has served a useful purpose. But if the group instead follows the most confident or highest-status agent, the society may converge on an incorrect answer even though the correct information was already present among its members. In this case, interaction has not simply failed to improve reasoning; it has changed how the information available to the society is used. We therefore see two different forms of belief change. The first is driven by \textit{epistemic information}: an agent changes its answer because another agent provides useful evidence, identifies an error, or presents a stronger argument. The second is driven by \textit{social influence}: an agent changes its answer because of properties of the source, such as confidence, reputation, authority, or social position. These two processes can produce the same observable outcome---an agent changing its answer---but they have very different implications for collective reasoning.

This distinction also changes how we should think about hierarchy in multi-agent systems. Hierarchy is not necessarily harmful. If the agents that receive greater social influence are also the agents that are more reliable, hierarchy may help a society coordinate efficiently and avoid unnecessary disagreement. The problem arises when social influence and epistemic competence become misaligned. A highly influential but unreliable agent can then affect many other agents, while a highly capable but socially peripheral agent may struggle to have its information heard. The resulting society may become more coordinated while becoming less accurate. The same issue applies to diversity. A society can contain diverse models and reasoning strategies at the beginning of an interaction, yet lose that diversity as agents repeatedly influence one another. Convergence toward a correct answer represents successful information sharing. Convergence toward an incorrect answer represents something very different: useful disagreement has been lost without the underlying uncertainty being resolved. Thus, the presence of diversity in a multi-agent system does not by itself guarantee that the society will benefit from that diversity. What matters is whether the information carried by that diversity survives the social process of interaction. Our perspective is deliberately broader than studying conformity or consensus alone. Agreement between agents is not inherently good or bad. What matters is \textit{why} the agents agree and whether the resulting collective belief is supported by the information available to the society. A group that reaches the correct answer after resolving disagreement has behaved very differently from a group that reaches the same answer because its members simply follow a socially dominant source.

We study these dynamics through the lens of \textbf{social-epistemic dynamics}. Our central focus is when social interaction helps a society use the knowledge distributed across its agents and when social structure causes that knowledge to be ignored. We investigate how different forms of agent diversity interact with social influence and hierarchy, and whether socially influential agents are in fact the most epistemically reliable members of the society. We are particularly interested in the boundary between beneficial and harmful convergence: the point at which interaction stops helping agents combine complementary information and instead begins to suppress useful disagreement. This view motivates a systematic study of multi-agent systems as \textbf{diverse societies of interacting reasoners}. Rather than viewing agents only as independent predictors whose outputs are aggregated at the end, we consider the interaction process itself as an important part of collective reasoning. The central hypothesis of this work is that collective reliability depends not only on the capabilities and diversity of individual agents, but also on how social influence determines which information is trusted, propagated, and ultimately retained by the society.


\section{Related Work}
\label{sec:related_work}

Multi-agent reasoning has emerged as a promising approach for improving
language-model reasoning by allowing multiple agents to exchange solutions,
critique one another, and iteratively refine their answers. A foundational
line of work established multi-agent debate as an inference-time mechanism
for combining the reasoning of multiple language-model instances.
\citet{du2024multiagent} introduced multi-agent debate as a general framework
in which multiple language-model instances interact over several rounds,
showing improvements across reasoning and factuality tasks.
\citet{liu2025dmad} showed that assigning agents different reasoning
strategies can improve debate by encouraging them to explore distinct
solution paths rather than repeatedly following similar reasoning patterns.
Similarly, \citet{zhu2025symphony} demonstrated that heterogeneous
language-model agents can provide complementary reasoning patterns for
multi-agent planning. These studies establish diversity and interaction as
important ingredients for collective reasoning, but largely focus on whether
diversity improves task performance rather than on how that diversity changes
as agents repeatedly influence one another.

At the same time, recent work has raised important questions about whether
interaction itself is responsible for the gains observed in multi-agent
reasoning. \citet{choi2025debate} systematically disentangled inter-agent
debate from majority voting and found that much of the improvement commonly
attributed to multi-agent debate can be explained by aggregation alone.
Their analysis highlights that belief changes during interaction should be
studied carefully rather than treating collective improvement as evidence
that communication itself is beneficial. Our work builds on this perspective
by examining the factors that determine how an agent responds to another
agent's information. In particular, we distinguish changes driven by
task-relevant evidence from changes driven by properties of the information
source.

This question connects closely to recent findings on conformity and
sycophancy in interacting language models. \citet{pitre2025consensagent}
identify sycophancy as a major reliability problem in multi-agent discussion,
showing that agents can reinforce one another's responses instead of
critically engaging with disagreement. \citet{choi2025conformity} provide a
broader empirical study of group conformity in multi-agent systems, finding
that agents can increasingly align with numerically dominant groups or more
capable agents during repeated discussions. These results demonstrate that
interaction can systematically change agent beliefs, but they primarily
characterize the resulting conformity or agreement. Our focus is on the
mechanism behind such changes: whether agents update because they encounter
stronger evidence, or because the source of the information carries greater
confidence, reputation, authority, or social influence.

Recent work has also begun to study language models as interacting
populations rather than simply as collections of independent predictors.
\citet{ashery2025social} show that decentralized populations of language-model
agents can spontaneously develop shared social conventions and collective
biases through repeated interaction. Their findings suggest that
population-level behavior can emerge from local interactions even when the
corresponding behavior is not explicitly specified for individual agents.
This provides an important foundation for studying multi-agent systems from a
social perspective. Our work considers a complementary setting in which
agents are solving tasks with objective answers, allowing us to ask whether
the social structures that emerge during interaction remain aligned with
epistemic competence.

The relationship between social influence and expertise is particularly
important for collective reasoning. \citet{pappu2026experts} recently showed
that self-organizing language-model teams can fail to effectively leverage
their strongest expert, instead tending toward compromise between expert
and non-expert opinions. Their results indicate that identifying a capable
agent is not sufficient; a team must also appropriately weight that agent's
information during interaction. This observation is closely related to our
focus on social hierarchy, but we consider a broader question: whether the
influence received by an agent is aligned with its actual epistemic
reliability. A socially prominent agent may be highly competent, in which
case hierarchy can help coordination, or it may be unreliable, in which case
the same social structure can amplify collective error.

Taken together, existing work shows that multi-agent interaction can improve
reasoning, that heterogeneous agents can provide complementary information,
and that repeated interaction can produce conformity, sycophancy, and
emergent social structures. However, these phenomena have largely been
examined as separate aspects of multi-agent systems. We bring them together
through a social-epistemic perspective, treating a multi-agent system as a
society in which agents possess different sources of information and occupy
different positions within an interaction structure. Our central distinction
is between \emph{epistemic influence}, where an agent changes its belief
because of useful task-relevant information, and \emph{social influence},
where the same change is driven by properties of the information source or
its social position. This allows us to study not only whether diverse agents
converge, but whether they converge because the society has successfully
integrated distributed knowledge or because social influence has caused
useful disagreement to disappear. We provide an extended discussion of
recent work on heterogeneous agents, social dynamics, emergent societies,
coordination, and multi-agent reliability in
Appendix~\ref{app:extended_literature}, together with a positioning analysis
showing how these directions relate to the dimensions studied in our
framework.


\section{Methodology}
\label{sec:methodology}

\begin{figure*}[t]
    \centering
    \includegraphics[width=\textwidth]{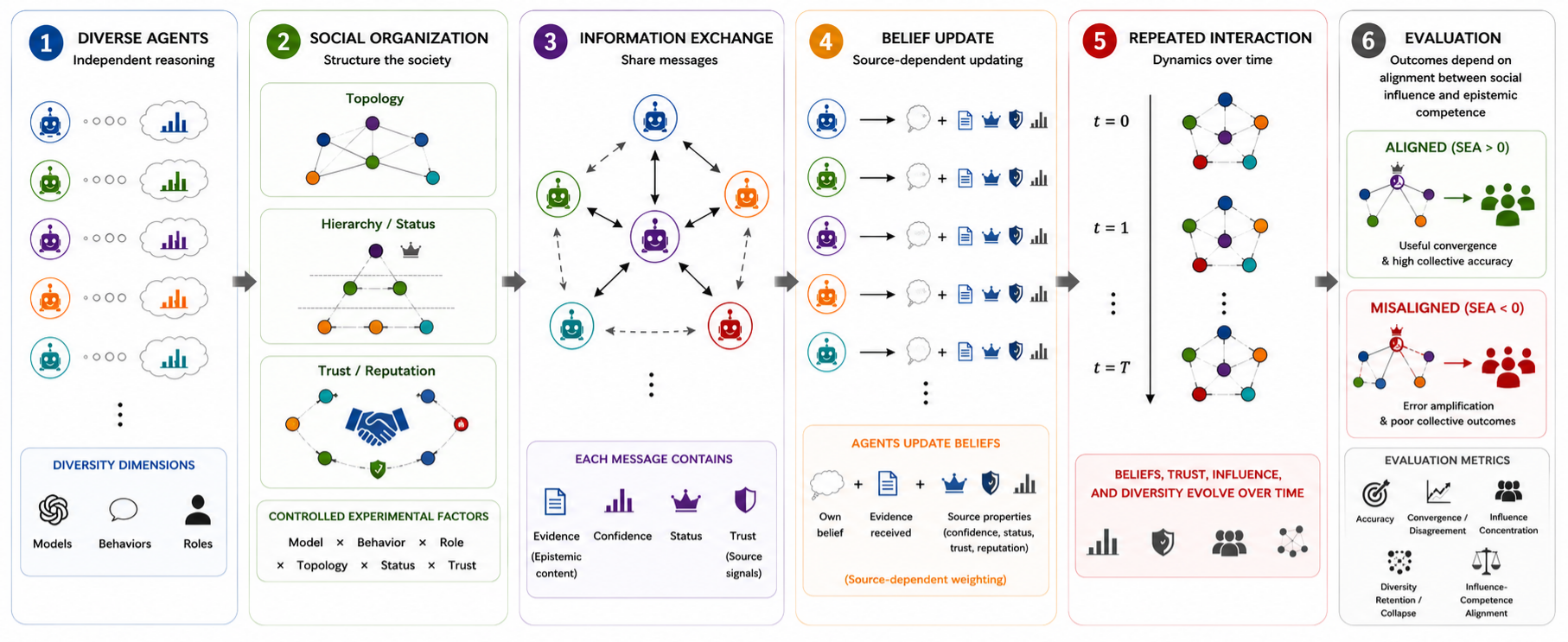}
    \caption{
    \textbf{End-to-end methodology overview.}
    Diverse agents begin with independent beliefs and are embedded in a
    social organization defined by model, behavioral, and role diversity,
    interaction topology, status, and trust. Agents exchange messages
    containing both epistemic content and social signals, after which
    beliefs, influence, trust, and disagreement evolve over repeated
    interaction. Depending on whether social influence aligns with
    epistemic competence, the same initial society can undergo useful
    convergence or collective error amplification. The illustrated scenario
    is illustrative; quantitative evaluation uses the benchmark tasks and
    controlled experimental conditions described in the paper.
    }
    \label{fig:methodology_overview}
\end{figure*}

We study a multi-agent system as a structured society rather than as a
collection of independent language-model calls. Agents enter the society
with different capabilities, behavioral policies, and roles, while their
collective behavior is shaped by the social organization through which
information is exchanged. As illustrated in Figure~\ref{fig:methodology_overview},
agents begin with independent beliefs, interact through a structured
information network, and repeatedly update their states in response to both
the content of received information and properties of its source. This
process can produce useful convergence when social influence tracks
epistemic competence, or collective error when socially influential
information is not epistemically reliable. The scenario shown in the figure
is illustrative; all quantitative evaluation uses the benchmark tasks and
controlled experimental conditions described later.

At a high level, we represent a society of $N$ agents as

\begin{equation}
\mathcal{S}
=
\left(
\mathcal{A},
\mathbf{G},
\mathbf{D},
\mathbf{H},
\mathbf{T}
\right),
\label{eq:method_society}
\end{equation}

where $\mathcal{A}=\{a_1,\ldots,a_N\}$ is the agent population,
$\mathbf{G}$ specifies the interaction topology, $\mathbf{D}$ specifies
agent diversity, $\mathbf{H}$ captures social status and influence, and
$\mathbf{T}$ represents trust or reputation. This decomposition allows
epistemic composition and social organization to be varied independently.
In particular, social status can contribute to an agent's social influence
without necessarily corresponding to its epistemic competence, allowing
competence-aligned and misaligned societies to be constructed
systematically.

Each agent first solves the task independently. At interaction round $t$,
its observable state is

\begin{equation}
S_i^t =
\left(
y_i^t,\,
c_i^t,\,
r_i,\,
h_i,\,
T_i^t
\right),
\label{eq:method_agent_state}
\end{equation}

where $y_i^t$ is the current answer or belief, $c_i^t$ is confidence,
$r_i$ is the assigned role, $h_i$ is social position, and $T_i^t$ is the
current trust or reputation state. We use only quantities observable from
model outputs and interaction histories and do not require access to hidden
chain-of-thought representations.

Agent diversity is controlled along three axes: model diversity, behavioral
diversity, and role diversity. Model diversity changes the underlying
language model, behavioral diversity changes the reasoning or interaction
policy, and role diversity changes the functional or social role assigned
to an agent. We encode these assignments as

\begin{equation}
\mathbf{D}
=
\left[
\mathbf{D}_M,\,
\mathbf{D}_B,\,
\mathbf{D}_R
\right],
\label{eq:method_diversity}
\end{equation}

where the three components correspond to model, behavioral, and role
assignments. Their detailed construction is provided in
Appendix~\ref{app:formal_society}.

Social organization determines how information propagates through the
population. Let

\begin{equation}
G_{ij}=1
\quad\Longleftrightarrow\quad
a_i \text{ can receive information from } a_j,
\label{eq:method_graph}
\end{equation}

where $\mathbf{G}\in\{0,1\}^{N\times N}$ defines the interaction topology.
We consider egalitarian, sparse, centralized, and hierarchical structures,
while independently manipulating whether social influence is aligned or
misaligned with epistemic competence. The resulting organization can also
restrict information flow through trust- or reputation-dependent access,
so that socially distant or low-trust sources may have limited influence
even when they are connected by the underlying interaction structure.
Detailed topology, social influence, and trust constructions are provided in
Appendix~\ref{app:social_structure}.

At each round, agent $i$ receives information from its accessible neighbors,

\begin{equation}
\mathcal{M}_i^t
=
\left\{
S_j^t : G_{ij}=1
\right\},
\label{eq:method_messages}
\end{equation}

and produces an updated state,

\begin{equation}
S_i^{t+1}
=
F_{\theta_i}
\left(
S_i^t,\,
\mathcal{M}_i^t,\,
r_i,\,
h_i,\,
T_i^t
\right),
\label{eq:method_update}
\end{equation}

where $F_{\theta_i}$ denotes the underlying model together with its
behavioral policy. An observed message contains both task-relevant
information and source-level social signals:

\begin{equation}
m_{ij}^t
=
\left(
e_{ij}^t,\,
c_j^t,\,
h_j,\,
T_{ij}^t
\right),
\label{eq:method_message}
\end{equation}

where $e_{ij}^t$ denotes task-relevant evidence, while the remaining
components capture source-level properties such as confidence, status, and
trust. Thus, a message is not treated as an undifferentiated answer:
epistemic content and source-level signals are represented separately,
allowing their contributions to subsequent belief change to be examined.

The resulting update is therefore source-dependent: the same task-relevant
evidence can receive different effective weight depending on the confidence,
status, or trust associated with its source. This provides the basis for
separating epistemic influence, arising from the informational content of a
message, from social influence, arising from properties of the source.

This decomposition defines our central distinction between
\emph{epistemic influence} and \emph{social influence}. Epistemic influence
occurs when belief change is attributable primarily to task-relevant
information; social influence occurs when the change is driven by
source-level properties while the underlying information is held fixed.
Our interventions vary these signals independently, allowing us to test
whether agents follow information because it is useful or because it comes
from a socially influential source.

Repeated interaction produces a society trajectory

\begin{equation}
\mathcal{S}^{0}
\rightarrow
\mathcal{S}^{1}
\rightarrow
\cdots
\rightarrow
\mathcal{S}^{T},
\label{eq:method_trajectory}
\end{equation}

through which beliefs, influence, trust, and disagreement may evolve.
We vary population size, topology, interaction rounds, diversity composition,
and the alignment between social status and epistemic competence. Canonical
conditions range from homogeneous and egalitarian societies to
heterogeneous societies with aligned, misaligned, and adversarial social
organization. The complete experimental matrix and paired evaluation
protocol are provided in Appendix~\ref{app:experimental_conditions}.

We evaluate both collective performance and the trajectory producing it.
In addition to individual and collective accuracy, we measure convergence,
retention of initial disagreement, influence concentration, diversity
retention or collapse, and alignment between social influence and epistemic
competence. We summarize the latter using

\begin{equation}
SEA
=
\rho\left(I,E\right),
\label{eq:method_sea}
\end{equation}

where $I_i$ is the measured social influence of agent $i$, $E_i$ is its
independently measured epistemic competence, and $\rho$ is their rank
correlation. Positive $SEA$ indicates that social influence preferentially
tracks epistemic competence, values near zero indicate weak alignment, and
negative $SEA$ indicates inverse alignment in which socially influential
agents tend to have lower epistemic competence. Full metric definitions,
statistical procedures, and implementation details are provided in
Appendix~\ref{app:evaluation_implementation}.

\section{Experimental Setup}
\label{sec:experimental_setup}

We evaluate the framework using heterogeneous open-weight language models
from the Qwen, Llama, Gemma, and Phi families across four core benchmarks:
GSM8K, MMLU-Pro, GPQA, and TruthfulQA. ALFWorld is additionally used as a
targeted interactive validation to test whether the observed
social-epistemic dynamics extend beyond static question answering. All
model parameters remain frozen during evaluation. For each controlled
condition, agents first solve matched task instances independently and then
participate in repeated interaction under the specified social
organization. The same task instances and initial responses are reused
across paired conditions whenever possible, allowing differences in
collective behavior to be attributed to the manipulated properties of the
society rather than to changes in task sampling or initial reasoning.
Complete model assignments, dataset construction and preprocessing,
society definitions, behavioral policies, network topologies, population
sizes, interaction schedules, controls, and experimental batches are
provided in Appendix~\ref{app:experimental_conditions}.

\section{Results and Discussions}
We organize the main results around three complementary findings that
progress from population-level effects to the underlying social-epistemic
mechanism. First, we examine how social organization determines collective
reliability across the canonical societies. Second, we characterize how
different social structures shape the evolution of beliefs, influence, and
epistemic diversity over repeated interaction. Third, we isolate the
mechanism underlying socially induced error by separating epistemic content
from source-level social signals and evaluating an evidence-gated
intervention. These results provide the main empirical picture; detailed
society-level comparisons, dataset and model breakdowns, topology and
population-size analyses, controls, sensitivity experiments, and statistical
analyses are provided in Appendix~\ref{app:experimental_results}.

\subsection{Social Organization Determines Collective Reliability}

We evaluate whether agent interaction improves or degrades collective performance across seven canonical society configurations ($S_0$--$S_6$) spanning static reasoning and interactive decision-making environments. Across all benchmarks, the presence of model or behavioral diversity alone is insufficient to guarantee collective intelligence; for instance, introducing unstructured combined heterogeneity ($S_3$) produces modest accuracy gains of +0.5 percentage points on GSM8K and +2.0 points on MMLU-Pro, while retaining only 0.35 of the initial epistemic diversity. Instead, the primary macro determinant of collective accuracy ($Acc_{\text{coll}}$) and error amplification ($\Delta Acc = Acc_T - Acc_0$) is \textbf{Social-Epistemic Alignment} ($SEA = \rho(I,E)$)---the degree to which social influence correlates with independently measured epistemic competence. When social influence is structurally aligned with epistemic competence ($S_4$), societies achieve substantial performance gains, including improvements of +3.5 percentage points on GSM8K, +5.0 points on MMLU-Pro, and +3.9 points on ALFWorld. Conversely, when influence is inversely aligned with competence ($S_6$), interaction causes systematic performance degradation, with decreases of -2.8 percentage points on GSM8K, -5.4 points on MMLU-Pro, and -1.8 points on ALFWorld. This relationship between social-epistemic alignment and collective reliability is not restricted to static evaluation: the competence-aligned and adversarial conditions exhibit corresponding effects in the interactive ALFWorld environment, where $S_4$ improves performance by +3.9 percentage points while $S_6$ decreases performance by -1.8 points. Table~\ref{tab:macro_results} provides a comprehensive map of these empirical outcomes across the canonical conditions.

\begin{table}[ht]
\centering
\caption{Master Empirical Summary Across Canonical Society Configurations. Summary of initial-to-final collective accuracy ($\Delta Acc$), Social-Epistemic Alignment ($SEA$), Influence Concentration ($IC$), and Diversity Retention ($DR$) across experimental batches. Statistical significance computed via paired bootstrap resampling against independent baselines. Point estimates are summarized here for compactness; paired task-level bootstrap 95\% confidence intervals are reported in the detailed benchmark-level analysis in Table 6.}
\label{tab:macro_results}
\resizebox{\textwidth}{!}{%
\begin{tabular}{@{}ll p{5.5cm} ccccccc@{}}
\toprule
\textbf{Batch} & \textbf{Society Condition} & \textbf{Manipulated Structural Property} & \textbf{GSM8K} & \textbf{MMLU-Pro} & \textbf{GPQA} & \textbf{ALFWorld} & \textbf{Mean} & \textbf{Mean} & \textbf{Mean} \\
& & & \textbf{($\Delta Acc$)} & \textbf{($\Delta Acc$)} & \textbf{($\Delta Acc$)} & \textbf{($\Delta Acc$)} & \textbf{$SEA$} & \textbf{$IC$} & \textbf{$DR$} \\
\midrule
\textbf{Baseline} & $S_0$: Homogeneous & Egalitarian baseline, identical models/policies & +0.0 pp & +0.0 pp & +0.0 pp & +0.0 pp & 0.00 & 0.25 & 1.00 \\
\addlinespace
\textbf{Diversity} & $S_1$: Model & Heterogeneous model capabilities ($D_M$) & +0.2 pp & +1.0 pp & -0.5 pp & +0.4 pp & +0.12 & 0.28 & 0.65 \\
& $S_2$: Behavioral & Heterogeneous reasoning policies ($D_B$) & +0.8 pp & -0.4 pp & +1.0 pp & +0.7 pp & +0.15 & 0.31 & 0.58 \\
& $S_3$: Combined & Unstructured model + behavioral heterogeneity & +0.5 pp & +2.0 pp & +0.2 pp & +0.8 pp & +0.18 & 0.35 & 0.35 \\
\addlinespace
\textbf{Alignment} & $S_4$: Aligned & Influence concentrated on competent nodes ($SEA > 0$) & +3.5 pp & +5.0 pp & +0.1 pp & +3.9 pp & +0.58 $\pm$ 0.07 & 0.68 & 0.42 \\
& $S_5$: Misaligned & Influence decoupled from competence ($SEA \approx 0$) & -1.1 pp & -1.9 pp & +0.3 pp & -0.8 pp & +0.04 $\pm$ 0.09 & 0.55 & 0.28 \\
& $S_6$: Adversarial & Influence concentrated on low-competence nodes ($SEA < 0$) & -2.8 pp & -5.4 pp & -0.8 pp & -1.8 pp & -0.51 $\pm$ 0.08 & 0.74 & 0.15 \\
\bottomrule
\multicolumn{10}{l}{\footnotesize \* $p < 0.05$, \*\* $p < 0.01$ relative to independent un-communicating baseline.}
\end{tabular}%
}
\end{table}

\subsection{Social Organization Shapes Collective Trajectories}
\label{sec:result_social_organization}

\begin{figure*}[t]
    \centering
    \includegraphics[width=\textwidth]{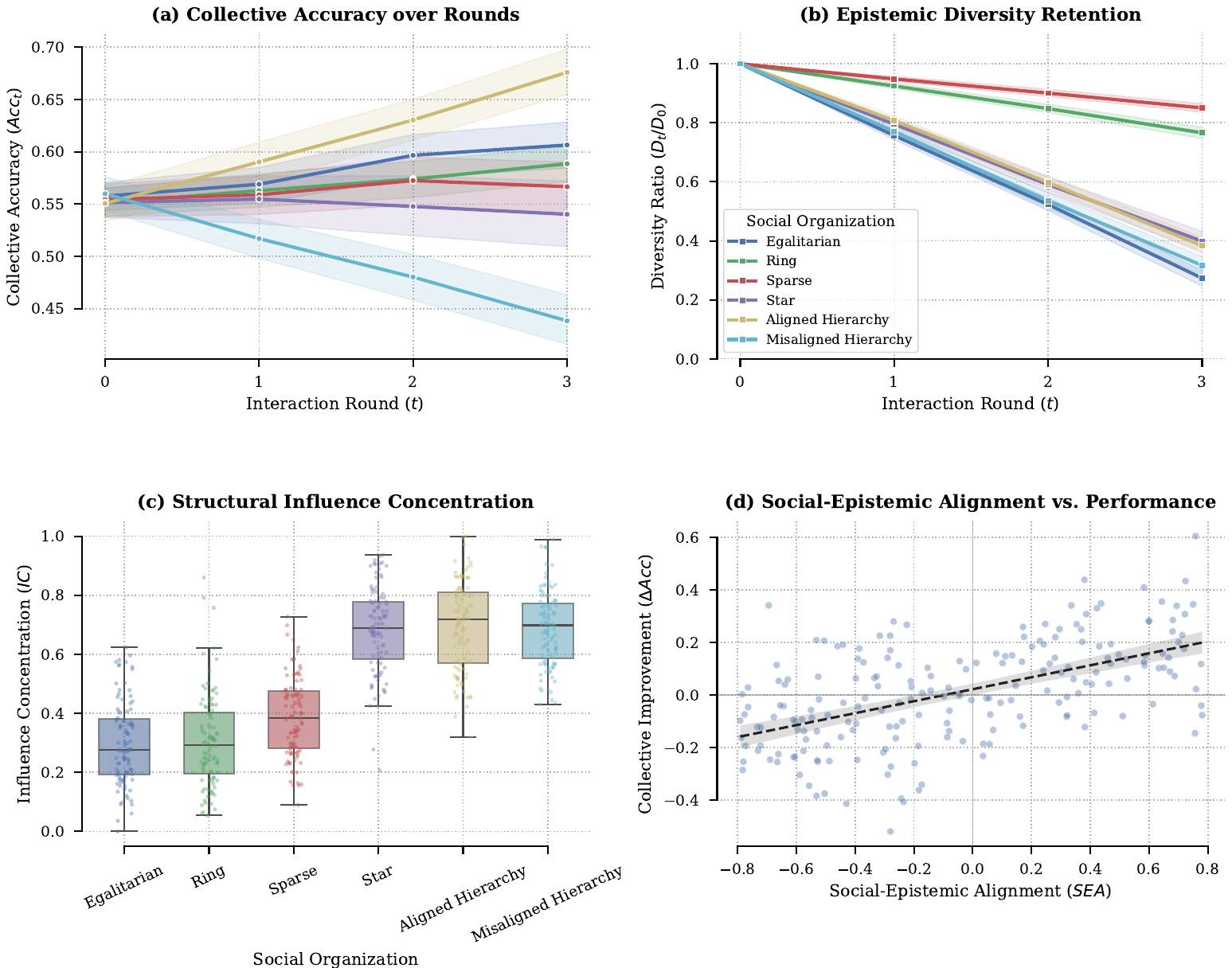}
    \caption{
    \textbf{Social organization shapes collective epistemic trajectories.}
    (a) Collective accuracy over repeated interaction rounds.
    (b) Retention of epistemic diversity relative to the initial society.
    (c) Concentration of social influence under different social
    organizations. (d) Association between social-epistemic alignment
    ($SEA$) and collective improvement relative to independent reasoning.
    Aligned and misaligned hierarchies use the same hierarchical
    organization but differ in whether social influence is aligned with
    independently measured epistemic competence.
    }
    \label{fig:social_organization_results}
\end{figure*}

Social organization substantially changes how collective reasoning evolves
over repeated interaction (Figure~\ref{fig:social_organization_results}).
The competence-aligned hierarchical society exhibits the strongest improvement
in collective accuracy, whereas competence-misaligned hierarchy progressively
degrades performance. Egalitarian, ring, and sparse organizations produce more
moderate changes, indicating that interaction alone does not guarantee either
improvement or failure.

The trajectories of epistemic diversity reveal a complementary effect:
interaction progressively reduces disagreement, but the rate of diversity
loss depends on the social organization. Thus, convergence should not be
interpreted as collective improvement by itself. Influence concentration
also increases as social organization becomes more centralized, yet the
aligned and misaligned hierarchical conditions can exhibit similarly
concentrated influence while producing opposite performance outcomes. This
indicates that influence concentration captures how strongly influence is
concentrated within the society, but does not by itself determine whether
that influence is beneficial.

This distinction is reflected in the positive association between
social-epistemic alignment ($SEA$) and collective improvement. Societies in
which social influence is more closely aligned with independently measured
epistemic competence tend to improve more relative to independent reasoning,
although substantial variation remains across individual experimental
instances. Overall, the results indicate that collective reliability depends
not simply on the presence or strength of social influence, but on whether the
social organization preferentially amplifies epistemically reliable
perspectives. Extended analyses across additional societies, topologies,
interaction depths, and controls are provided in
Appendix~\ref{app:experimental_results}.

\subsection{Social Influence Can Override Evidence}
\label{sec:result_social_override}

\begin{figure*}[t]
    \centering
    \includegraphics[width=\textwidth]{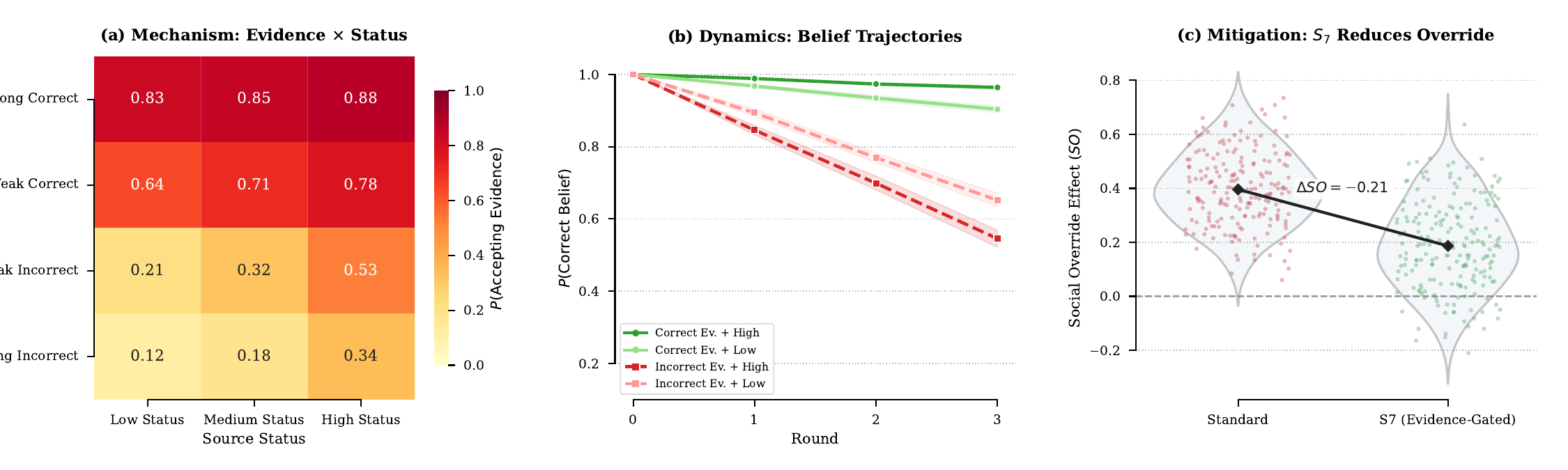}
    \caption{
    \textbf{Social influence can override evidence, but evidence-gated
    updating reduces the effect.}
    (a) Probability of accepting evidence as a function of evidence quality
    and source status. Higher-status sources receive greater weight,
    particularly when the evidence is weak or incorrect. (b) Repeated
    interaction causes the effect to accumulate in belief trajectories,
    with incorrect evidence from high-status sources producing progressively
    larger degradation. (c) Evidence-gated updating ($S_7$) substantially
    reduces the social-override effect relative to standard updating.
    }
    \label{fig:result_5_3}
\end{figure*}

The preceding result showed that social organization can substantially alter collective reasoning trajectories. We next examine the mechanism underlying this effect: whether agents respond primarily to the epistemic content of information or to properties of the source providing it. Figure~\ref{fig:result_5_3} summarizes this mechanism, its effect over repeated interaction, and the corresponding mitigation. Figure~\ref{fig:result_5_3}(a) shows that evidence acceptance depends jointly on evidence quality and source status. Strong correct evidence is generally accepted across status levels, whereas weak or incorrect evidence receives substantially greater weight when presented by high-status agents. Thus, source-level properties can alter belief updating even when the underlying epistemic content is held fixed, providing a direct separation between epistemic and social influence. This source-dependent weighting also propagates across interaction rounds, as shown in Figure~\ref{fig:result_5_3}(b). Beliefs exposed to correct evidence remain comparatively stable, while incorrect evidence from high-status sources produces progressively larger degradation. This indicates that social influence can transform a local source-weighting bias into a collective belief trajectory rather than merely affecting a single update. Finally, Figure~\ref{fig:result_5_3}(c) compares standard updating with the evidence-gated intervention $S_7$. The social-override effect decreases substantially under $S_7$, while the remaining overlap between distributions shows that the intervention reduces rather than eliminates socially induced updating. Taken together, these results identify a concrete failure mechanism in which social status causes agents to overweight unreliable information, with this bias accumulating through repeated interaction. Evidence-gated updating provides a simple intervention for partially decoupling social influence from epistemic reliability.

\section{Limitations}
Our study focuses on controlled multi-agent societies built from contemporary
language models and therefore does not exhaust the space of possible agent
architectures, interaction protocols, or real-world social environments.
Although we evaluate multiple models, benchmarks, network structures, and an
interactive environment, the experiments remain primarily task- and
simulation-based. The proposed evidence-gated intervention is also evaluated
as a targeted mitigation rather than as a comprehensive solution to social
influence failures. Future work should examine longer-horizon interactions,
richer forms of social organization, and settings in which agent competence
and social status evolve dynamically over time.


\section{Conclusion}
\label{sec:conclusion}

We studied multi-agent reasoning as a social-epistemic process, focusing on
when interaction allows the knowledge distributed across a diverse society
to become a collective advantage and when it instead causes that knowledge
to be lost. Our results suggest that diversity creates complementary
epistemic resources, but social organization determines which of these
resources are heard, trusted, and propagated. When influence is aligned with
epistemic competence, interaction can turn disagreement into collective
improvement; when misaligned, the same process can amplify collective error.
Notably, adversarial influence can reverse these gains, producing a $5.4$
percentage-point degradation on MMLU-Pro. The broader evidence therefore
suggests that diversity should not be treated as an end in itself: its value
depends on whether the social process preserves and appropriately amplifies
the information it creates.

Our experiments further show that social influence changes not only which
answer ultimately prevails, but how agents interpret evidence during
interaction. Higher-status sources receive greater weight even when their
evidence is weak or incorrect, allowing local source-weighting biases to
compound over repeated interaction, while influence directed toward
epistemically reliable agents can propagate useful corrections. This
suggests that hierarchy, influence, and convergence are not inherently
beneficial or harmful; their consequences depend on the epistemic quality
of what they amplify. Reliable multi-agent systems should therefore
preserve useful disagreement long enough for its epistemic value to be
evaluated, while keeping social influence grounded in evidence. Our
evidence-gated intervention supports this direction by reducing socially
induced updates when they conflict with available evidence. More broadly,
this reframes social structure from a communication detail into a component
of the reasoning architecture itself: how agents are connected and
influenced determines which capabilities become collective knowledge.
Ultimately, collective intelligence depends not only on what agents know,
but on the social process determining whose knowledge the society allows
to matter.

%
% ...

\newpage
\bibliographystyle{plainnat}
\bibliography{references}

\newpage
\input{checklist}

\newpage
\appendix
\section*{Appendix}

\appendix


\appendix

\section{Extended Literature Review and Positioning}
\label{app:extended_literature}

This appendix provides additional context on recent work surrounding diverse,
interacting, and socially structured language-model agents. The main paper
discusses the core literature needed to motivate our central research
question. Here, we broaden the discussion to include recent work on social
influence, collective belief dynamics, heterogeneous agents, coordination,
emergent societies, multi-agent reliability, and resource-efficient agent
systems. We focus primarily on recent work from 2025--2026, together with
selected earlier work that directly motivates dimensions of our setting. The
papers discussed here complement the main Related Work section rather than
repeating its central references.


\subsection{Social Influence and Collective Dynamics}
\label{app:social_dynamics}

Recent work has increasingly shown that the behavior of interacting language
models cannot always be understood from their individual capabilities alone.
\citet{ko2026social} systematically study social conformity, perceived
expertise, dominant-speaker effects, and rhetorical persuasion in LLM
collectives. Their experiments show that increasing social pressure can
reduce the accuracy of an agent even when the influencing agents are capable
reasoners. This provides a direct connection to our treatment of social
influence, while our setting additionally considers how such effects interact
with model, behavioral, and role diversity.

\citet{choi2025conformity} provide another empirical study of group conformity
in multi-agent LLM interactions. Across thousands of simulated debates, they
find that agents can shift toward numerically dominant groups and more
intelligent agents. Their findings suggest that both group structure and
perceived capability can affect the direction of collective belief updates.
Our work builds on this observation by separating actual epistemic competence
from social position and by explicitly studying cases where the two are
aligned or misaligned.

Recent work has also connected collective behavior to the structure of the
interaction network. \citet{liu2026snla} introduce Social Networks of LLM
Agents (SNLA), which models influence as a function of network position and
the attention that agents allocate to their peers. Their analysis shows that
narrow attention can lead to herding and limit the effective amount of
information aggregated by a population, even as the nominal population
grows. This is particularly relevant to our manipulation of interaction
topology and population size.

\citet{abdurahman2026trait} study how agents infer interaction-partner
characteristics such as warmth and competence from interaction histories.
Their results show that explicit representations of partner traits can
improve coordination. This provides a useful connection to our treatment of
trust, competence, and reputation. However, our goal is not primarily to
improve coordination through trait inference; instead, we study how such
social judgments can affect which information agents choose to trust and
follow.

\citet{ashery2025social} provide evidence that repeated interaction can
produce social conventions and collective biases even when such conventions
are not explicitly specified at the individual level. Their experiments
demonstrate the emergence of population-level conventions and the ability of
committed minorities to alter established conventions. Our setting is
complementary: we study societies operating on tasks with objectively
evaluable answers, allowing social dynamics to be connected directly to
epistemic reliability.


\subsection{Diversity, Heterogeneity, and Coordination}
\label{app:diversity_coordination}

A second line of work studies how diversity across agents affects collective
reasoning and coordination. \citet{zhu2026demystifying} show that multi-agent
debate can benefit from diversity in initial viewpoints and from explicit
confidence communication. Their results suggest that homogeneous agents
following similar update rules may fail to obtain the full benefits of
interaction. This finding is closely related to our distinction between
initial epistemic diversity and diversity that survives repeated interaction.

\citet{liu2025dmad} similarly show that assigning different reasoning
strategies to agents can break fixed reasoning patterns and improve
multi-agent debate. SYMPHONY~\cite{zhu2025symphony} extends this idea to
planning by combining heterogeneous language models to increase exploration
diversity. These studies motivate our use of model and behavioral diversity,
but primarily treat diversity as a mechanism for improving collective
performance. We instead study whether such diversity remains effective after
agents are exposed to social influence.

\citet{chen2025ioa} introduce the Internet of Agents, a framework for
connecting heterogeneous agents through dynamic teaming and communication.
Their work demonstrates that agents originating from different model
families can collaborate through a shared interaction architecture.
Hetero-Designer~\cite{zhang2026hetero} further investigates automated design
of heterogeneous multi-agent systems by selecting different models and
roles. These approaches support the practical importance of heterogeneous
populations, while our experiments focus on the subsequent dynamics of those
populations rather than on optimizing their initial composition.

\citet{li2025semdiv} study semantically diverse teammates and show that
agents with different coordination behaviors can improve generalization to
unseen partners. Their results motivate our distinction between model
diversity and behavioral diversity: two agents may share an underlying model
while still behaving differently during interaction.

The organization of interaction is another important source of variation.
\citet{yang2025agentnet} investigate decentralized coordination in which
agents can dynamically specialize and change their connections.
\citet{dang2025evolving} study evolving orchestration, where the organization
of multi-agent collaboration changes in response to task state. These works
demonstrate that topology and organizational structure can be computationally
meaningful. Our framework instead treats topology and social organization as
controlled experimental variables so that their effects on collective belief
dynamics can be isolated.

\citet{xie2025econ} provide a complementary game-theoretic perspective by
modeling multi-agent reasoning under incomplete information and beliefs about
other agents. \citet{jha2025cec} study cross-environment cooperation and show
that agents can learn coordination patterns that generalize to new partners
and tasks. Together, these works highlight the importance of modeling
interaction as a process involving beliefs about other agents rather than
merely aggregating independent model outputs.


\subsection{Emergent Societies, Information, and Reliability}
\label{app:societies_reliability}

Beyond task-oriented multi-agent frameworks, recent research increasingly
treats LLM agents as populations capable of developing persistent social
structures. \citet{hashemi2026collective} analyze 7 million posts and
interactions among approximately 32,000 LLM agents on an LLM-driven social
platform. They report patterns of homophily and social influence, together
with population-level polarization and other collective effects. This
large-scale setting provides evidence that social structures can emerge from
repeated interactions between language-model agents.

\citet{willis2026collective} study collective behavior among hundreds of LLM
agents in social-dilemma environments, examining how population-level
outcomes emerge from repeated interactions. \citet{arruda2026cooperation}
similarly compare networked cooperation among LLM-agent populations with
human interaction data. These works emphasize that aggregate population
behavior can differ substantially from the behavior of individual agents.
Our framework complements this line of work by connecting population
dynamics to objectively measurable epistemic performance.

Information distribution provides another important connection.
\citet{liu2024iasym} study collaborative tasks under information asymmetry,
where different agents possess different pieces of information and
communication is required to reconstruct a shared problem state. Their
results motivate the idea that the value of a society may lie in its
distributed information rather than in the capability of any single member.

\citet{cemri2025masta} introduce MAST-Data and a taxonomy of failure modes in
multi-agent LLM systems. Their analysis identifies failures arising from
system design, inter-agent misalignment, and task verification, showing that
interaction can introduce failure modes that are not visible when agents are
evaluated independently.

\citet{huang2025resilience} study the resilience of multi-agent systems in
the presence of faulty agents and compare alternative communication
structures. Their results show that the organization of the agent population
can substantially affect how errors propagate. Our setting is related but
does not require explicitly faulty agents: instead, we ask whether an
agent's social influence is appropriately aligned with its actual epistemic
competence.

\citet{keshmirian2025utilitarian} study multi-agent moral reasoning and show
that group deliberation can systematically change judgments relative to
isolated agents. This provides additional evidence that collective
interaction can alter decisions even when the underlying models and tasks
remain unchanged.

\citet{hua2025shapleycoop} investigate cooperation among self-interested LLM
agents and use contribution-aware mechanisms to encourage collaboration.
Their work demonstrates that information about an agent's contribution can
become part of the social coordination process. This is related to our
treatment of status and influence, although our focus is on whether socially
influential agents are actually epistemically reliable.


\subsection{Small Language Models and Broader Multi-Agent Systems}
\label{app:slm_mas}

Small language models provide another meaningful source of heterogeneity for
agent societies. \citet{wang2026slm} study open-source models below 10B
parameters across direct prompting, tool-augmented single-agent settings,
and routing-based multi-agent systems. The study finds that structured
agentic approaches can improve small-model effectiveness, while multi-agent
routing can introduce additional coordination overhead. This motivates our
use of smaller models as members of heterogeneous societies rather than
treating them only as cheaper substitutes for larger models.

\citet{liu2025lessons} study collaborative learning among code language
models and demonstrate that agents can exchange useful information through
interaction. Their findings provide an additional motivation for including
capability differences within an interacting population.

\citet{lamalfa2025mas} argue that many current LLM-based multi-agent systems
do not fully engage with concepts from classical multi-agent systems,
particularly social agency, structured environments, communication
protocols, and emergent behavior. This perspective motivates our explicit
treatment of the interaction structure and social organization of a
population rather than viewing a multi-agent system simply as a collection
of independent LLM calls.

Collectively, the literature above establishes several important components
of our setting: social influence can distort decisions; diversity can improve
collective reasoning; network structure can alter information flow;
heterogeneous models and roles can provide complementary capabilities;
repeated interaction can generate population-level social structures; and
multi-agent coordination can introduce new failure modes. Our work connects
these observations through a common social-epistemic perspective.

In particular, we distinguish between \emph{epistemic influence}, arising
from task-relevant information, and \emph{social influence}, arising from
properties of the information source such as confidence, authority, status,
or reputation. This allows us to study whether social influence remains
aligned with epistemic competence and whether initially diverse beliefs are
preserved or collapse during repeated interaction. The experimental framework
introduced in Section~\ref{sec:methodology} therefore varies model,
behavioral, and role diversity together with social hierarchy, interaction
topology, and population size.
Table~\ref{tab:extended_literature} summarizes the additional literature
discussed in this appendix and its relationship to the present study.


\begin{table*}[t]
\centering
\scriptsize
\setlength{\tabcolsep}{3pt}
\renewcommand{\arraystretch}{1.15}

\caption{
Positioning of recent related work along the dimensions relevant to our
social-epistemic framework. A $\checkmark$ indicates that an aspect is a
central component of the cited work, while $\sim$ indicates a related but
secondary treatment. Our work jointly studies all six dimensions, with
particular emphasis on whether social influence remains aligned with
epistemic competence as heterogeneous agent societies evolve through
repeated interaction.
}
\label{tab:extended_literature}

\begin{tabularx}{\textwidth}{
>{\raggedright\arraybackslash}p{0.28\textwidth}
*{6}{>{\centering\arraybackslash}X}
}

\toprule

\textbf{Work}
&
\makecell{\textbf{Interaction}}
&
\makecell{\textbf{Diversity}}
&
\makecell{\textbf{Social}\\\textbf{Structure}}
&
\makecell{\textbf{Epistemic}\\\textbf{Competence}}
&
\makecell{\textbf{Dynamic}\\\textbf{Society}}
&
\makecell{\textbf{Epistemic}\\\textbf{Reliability}}
\\

\midrule

Ko et al.~\cite{ko2026social}
& $\checkmark$ & $\sim$ & $\checkmark$ & $\checkmark$ & $\checkmark$ & $\checkmark$ \\

Choi et al.~\cite{choi2025conformity}
& $\checkmark$ & $\sim$ & $\checkmark$ & $\checkmark$ & $\checkmark$ & $\checkmark$ \\

Liu et al.~\cite{liu2026snla}
& $\checkmark$ & -- & $\checkmark$ & $\sim$ & $\checkmark$ & $\checkmark$ \\

Abdurahman et al.~\cite{abdurahman2026trait}
& $\checkmark$ & -- & $\checkmark$ & $\checkmark$ & $\checkmark$ & $\sim$ \\

Ashery et al.~\cite{ashery2025social}
& $\checkmark$ & $\sim$ & $\checkmark$ & -- & $\checkmark$ & $\sim$ \\

Zhu et al.~\cite{zhu2026demystifying}
& $\checkmark$ & $\checkmark$ & $\sim$ & $\checkmark$ & $\checkmark$ & $\checkmark$ \\

Chen et al.~\cite{chen2025ioa}
& $\checkmark$ & $\checkmark$ & $\checkmark$ & $\sim$ & $\sim$ & -- \\

Li et al.~\cite{li2025semdiv}
& $\checkmark$ & $\checkmark$ & -- & $\sim$ & $\sim$ & $\sim$ \\

Zhang et al.~\cite{zhang2026hetero}
& $\checkmark$ & $\checkmark$ & $\checkmark$ & $\sim$ & $\sim$ & -- \\

Yang et al.~\cite{yang2025agentnet}
& $\checkmark$ & $\checkmark$ & $\checkmark$ & -- & $\checkmark$ & $\sim$ \\

Dang et al.~\cite{dang2025evolving}
& $\checkmark$ & $\sim$ & $\checkmark$ & -- & $\checkmark$ & $\sim$ \\

Xie et al.~\cite{xie2025econ}
& $\checkmark$ & -- & $\checkmark$ & $\checkmark$ & $\sim$ & $\checkmark$ \\

Jha et al.~\cite{jha2025cec}
& $\checkmark$ & $\sim$ & $\sim$ & -- & $\checkmark$ & $\sim$ \\

Hashemi \& Macy~\cite{hashemi2026collective}
& $\checkmark$ & $\sim$ & $\checkmark$ & $\sim$ & $\checkmark$ & $\sim$ \\

Willis et al.~\cite{willis2026collective}
& $\checkmark$ & $\sim$ & $\checkmark$ & -- & $\checkmark$ & $\sim$ \\

Cemri et al.~\cite{cemri2025masta}
& $\checkmark$ & $\sim$ & $\sim$ & $\sim$ & $\sim$ & $\checkmark$ \\

Huang et al.~\cite{huang2025resilience}
& $\checkmark$ & $\sim$ & $\checkmark$ & $\checkmark$ & $\sim$ & $\checkmark$ \\

Keshmirian et al.~\cite{keshmirian2025utilitarian}
& $\checkmark$ & $\sim$ & $\sim$ & -- & $\checkmark$ & $\checkmark$ \\

Hua et al.~\cite{hua2025shapleycoop}
& $\checkmark$ & $\sim$ & $\checkmark$ & $\sim$ & $\checkmark$ & $\sim$ \\

Wang~\cite{wang2026slm}
& $\checkmark$ & $\checkmark$ & $\sim$ & $\checkmark$ & $\sim$ & $\sim$ \\

Liu et al.~\cite{liu2025lessons}
& $\checkmark$ & $\checkmark$ & $\sim$ & $\sim$ & $\checkmark$ & $\sim$ \\

La Malfa et al.~\cite{lamalfa2025mas}
& $\checkmark$ & $\sim$ & $\checkmark$ & $\sim$ & $\checkmark$ & $\sim$ \\

\midrule

\textbf{Our work}
& $\checkmark$
& $\checkmark$
& $\checkmark$
& $\checkmark$
& $\checkmark$
& $\checkmark$
\\

\bottomrule

\end{tabularx}

\vspace{1mm}

\begin{flushleft}
\scriptsize
\textbf{Legend:}
$\checkmark$ = central focus;
$\sim$ = related but not a primary focus;
-- = not explicitly studied.
\end{flushleft}

\end{table*}



\section{Extended Methodology}
\label{app:formal_methodology}

This appendix provides the formal specification of the agent societies,
diversity assignments, social structure, interaction process, and
social-epistemic dynamics used in the main experiments. The notation
expands the compact description given in Section~\ref{sec:methodology}.
The organization follows the methodological flow illustrated in
Figure~\ref{fig:methodology_overview}: agents begin with diverse
capabilities and roles, are embedded in a social structure, exchange
information containing both epistemic and social signals, and evolve
through repeated interaction.

\subsection{Formal Society and Diversity}
\label{app:formal_society}

We represent a society as

\begin{equation}
\mathcal{S}
=
\left(
\mathcal{A},
\mathbf{G},
\mathbf{D},
\mathbf{H},
\mathbf{T}
\right),
\label{eq:app_society}
\end{equation}

where $\mathcal{A}$ is the set of agents, $\mathbf{G}$ is the interaction
topology, $\mathbf{D}$ describes agent diversity, $\mathbf{H}$ represents
social status and influence, and $\mathbf{T}$ represents pairwise trust or
reputation.

For a society containing $N$ agents,

\begin{equation}
\mathcal{A}
=
\{a_1,a_2,\ldots,a_N\}.
\label{eq:app_agent_set}
\end{equation}

Each agent is associated with a model identity, behavioral policy,
functional role, and social position. These properties remain fixed within
an experimental condition unless the corresponding experiment explicitly
studies their evolution.

The interaction topology is represented by

\begin{equation}
\mathbf{G}
\in
\{0,1\}^{N\times N},
\qquad
G_{ij}=1
\Longleftrightarrow
a_i \text{ can receive information from } a_j,
\label{eq:app_graph}
\end{equation}

with self-interaction excluded by setting $G_{ii}=0$.

At interaction round $t$, agent $a_i$ has the observable state

\begin{equation}
S_i^t
=
\left(
y_i^t,
c_i^t,
r_i,
h_i,
T_i^t
\right),
\label{eq:app_agent_state}
\end{equation}

where $y_i^t$ denotes the current answer or belief, $c_i^t$ denotes
reported confidence, $r_i$ denotes the behavioral or functional role,
$h_i$ denotes social position, and $T_i^t$ denotes the current trust or
reputation state.

We restrict the formal state to quantities observable from model outputs
and interaction histories and do not require access to hidden
chain-of-thought representations.

Before social interaction, every agent independently solves the task:

\begin{equation}
S_i^0
=
F_{\theta_i}^{(0)}(x).
\label{eq:app_initial_state}
\end{equation}

We introduce diversity along three controlled dimensions. Model diversity
assigns different language models to members of the same society.
Behavioral diversity assigns different interaction policies, including
exploration, verification, criticism, and conservative updating. Role
diversity assigns different functional roles within the society.

The model assignments are

\begin{equation}
\mathbf{D}_M
=
(d_{M,1},\ldots,d_{M,N}),
\qquad
d_{M,i}\in\{1,\ldots,K_M\},
\label{eq:app_model_diversity}
\end{equation}

behavioral assignments are

\begin{equation}
\mathbf{D}_B
=
(d_{B,1},\ldots,d_{B,N}),
\qquad
d_{B,i}\in\{1,\ldots,K_B\},
\label{eq:app_behavior_diversity}
\end{equation}

and role assignments are

\begin{equation}
\mathbf{D}_R
=
(d_{R,1},\ldots,d_{R,N}),
\qquad
d_{R,i}\in\{1,\ldots,K_R\}.
\label{eq:app_role_diversity}
\end{equation}

The combined diversity specification is

\begin{equation}
\mathbf{D}
=
\left[
\mathbf{D}_M,
\mathbf{D}_B,
\mathbf{D}_R
\right].
\label{eq:app_combined_diversity}
\end{equation}

This decomposition allows each source of diversity to be manipulated while
the others are held fixed.

\subsection{Social Structure and Interaction Dynamics}
\label{app:social_structure}

Social structure determines how information is exposed, weighted, and
propagated within the society. We model it through interaction topology,
social status or hierarchy, and trust or reputation.

The social influence structure is represented by

\begin{equation}
\mathbf{H}
\in
\mathbb{R}^{N\times N},
\label{eq:app_hierarchy_matrix}
\end{equation}

where $H_{ij}$ represents the social weight assigned to information
originating from agent $j$ when considered by agent $i$.

The aggregate influence of agent $j$ is

\begin{equation}
I_j
=
\sum_{i=1}^{N} H_{ij}.
\label{eq:app_agent_influence}
\end{equation}

Let $E_i$ denote independently measured epistemic competence. We
distinguish societies according to the alignment between influence and
competence:

\begin{equation}
SEA
=
\rho(I,E),
\label{eq:app_structural_sea}
\end{equation}

where $\rho$ denotes rank correlation across agents.

Aligned societies assign greater social influence to more reliable agents,
whereas misaligned societies assign influence independently of, or
inversely to, epistemic competence.

Trust and reputation are represented by

\begin{equation}
\mathbf{T}^{t}
=
[T_{ij}^{t}]_{i,j=1}^{N},
\qquad
T_{ij}^{t}\in[0,1].
\label{eq:app_trust_matrix}
\end{equation}

Trust may be initialized uniformly,

\begin{equation}
T_{ij}^{0}=T_0,
\label{eq:app_initial_trust}
\end{equation}

or according to the corresponding social condition. When dynamic trust is
enabled, updates depend only on observable interaction outcomes.

The topology $\mathbf{G}$ may represent fully connected, ring, star,
sparse, or hierarchical communication structures. Unless explicitly
studied, the topology remains fixed throughout a run.

At round $t$, agent $i$ receives information from its accessible neighbors:

\begin{equation}
\mathcal{M}_i^t
=
\left\{
S_j^t:
G_{ij}=1
\right\}.
\label{eq:app_message_set}
\end{equation}

The agent then updates its state according to

\begin{equation}
S_i^{t+1}
=
F_{\theta_i}
\left(
S_i^t,
\mathcal{M}_i^t,
r_i,
h_i,
T_i^t
\right).
\label{eq:app_formal_update}
\end{equation}

To separate epistemic and social influence, a received message is written
as

\begin{equation}
m_{ij}^t
=
\left(
e_{ij}^t,
c_j^t,
h_j,
T_{ij}^t
\right),
\label{eq:app_message_components}
\end{equation}

where $e_{ij}^t$ denotes task-relevant evidence and the remaining
components represent source-level signals.

Epistemic influence refers to belief change attributable primarily to the
information content. Social influence refers to change attributable
primarily to source-level properties while the underlying information is
held fixed. We operationalize this distinction through paired interventions
that vary social signals while preserving informational content.

The resulting society trajectory is

\begin{equation}
\mathcal{S}^{0}
\rightarrow
\mathcal{S}^{1}
\rightarrow
\cdots
\rightarrow
\mathcal{S}^{T}.
\label{eq:app_society_trajectory}
\end{equation}



\section{Extended Experimental Methodology}
\label{app:experimental_conditions}

This appendix specifies the experimental protocol used to instantiate and
evaluate the societies introduced in the main paper. We use the term
\emph{colony} to denote one instantiated population of agents solving the
same task under a specified social organization. A single task instance can
therefore be evaluated across multiple colonies, enabling paired
comparisons in which the task, ground truth, and initial independent
responses are held fixed while the social condition is varied.

The experimental design separates three layers. The \emph{task layer}
defines the benchmark instance and ground-truth answer. The \emph{agent
layer} determines model, behavioral policy, and role assignments. The
\emph{social layer} determines information availability, social status,
influence, reputation, and interaction dynamics. This separation allows
changes in collective behavior to be attributed to specific sources of
variation rather than to changes in the underlying task.


\subsection{Datasets and Task Construction}
\label{app:datasets}

We evaluate the framework on reasoning and knowledge tasks spanning
mathematical reasoning, broad academic knowledge, difficult expert
reasoning, factual reliability, and interactive decision making. The core
benchmarks are GSM8K, MMLU-Pro, GPQA, and TruthfulQA, while ALFWorld is used
as a targeted interactive validation. The benchmarks retain their original
semantic content and ground-truth definitions. The principal experimental
manipulations are therefore applied at the agent and social layers rather
than by rewriting benchmark questions.

\begin{table*}[t]
\centering
\small
\setlength{\tabcolsep}{4pt}
\renewcommand{\arraystretch}{1.12}
\caption{Benchmark coverage across the experimental program. The listed
use denotes the primary role of each benchmark; exact task subsets and
model assignments are recorded for each experimental run.}
\label{tab:dataset_coverage}
\begin{tabularx}{\textwidth}{
    l
    l
    >{\raggedright\arraybackslash}X
    l
    >{\raggedright\arraybackslash}X
}
\toprule
\textbf{Benchmark}
&
\textbf{Task Type}
&
\textbf{Primary Experimental Use}
&
\textbf{Model Coverage}
&
\textbf{Purpose}
\\
\midrule

GSM8K
&
Mathematical reasoning
&
Diversity, alignment, social-signal analysis
&
Qwen/Llama/Gemma/Phi
&
Objective reasoning with well-defined answers
\\

MMLU-Pro
&
Multi-domain reasoning
&
Diversity, topology, alignment
&
Qwen/Llama/Gemma/Phi
&
Tests generality across knowledge domains
\\

GPQA
&
Expert reasoning
&
Competence alignment, adversarial influence
&
Qwen/Llama/Gemma/Phi
&
Provides stronger variation in independent competence
\\

TruthfulQA
&
Factual reliability
&
Social influence, confidence
&
Compatible model subset
&
Tests settings where confidence and correctness can diverge
\\

ALFWorld
&
Interactive decision making
&
Targeted interactive validation
&
Compatible instruction models
&
Tests extension beyond static question answering
\\

\bottomrule
\end{tabularx}
\end{table*}

GSM8K provides mathematical reasoning problems with well-defined answers.
MMLU-Pro provides broad multi-domain reasoning tasks. GPQA provides
difficult expert-level questions where independent epistemic competence can
differ substantially. TruthfulQA provides a complementary factual
reliability setting in which agreement and confidence need not correspond
to correctness. ALFWorld provides a targeted interactive setting for
testing whether the observed social-epistemic dynamics extend beyond
static question answering.

For each benchmark, task instances are divided into fixed experimental
subsets before the social experiments are run. The same task instance is
reused across corresponding colonies whenever a paired comparison is
possible. Ground-truth answers and benchmark evaluation procedures remain
fixed across these comparisons.

We distinguish \emph{task construction} from \emph{social intervention}.
Benchmark questions are not rewritten merely to create different societies.
Instead, the same task is presented under different model compositions,
behavioral policies, roles, interaction topologies, and social conditions.
This prevents social effects from being confounded with changes in task
difficulty.

For content-preserving social-signal experiments, the task-relevant
information is held fixed while observable source-level properties are
manipulated. These properties include confidence, status, authority, or
reputation. The intervention therefore operates on the social context of a
response rather than on the benchmark question or ground truth.

No hidden chain-of-thought representation is required. All experimental
manipulations are defined over observable answers, confidence estimates,
response content, interaction histories, roles, and social metadata.


\subsection{Model Pool and Agent Assignment}
\label{app:model_pool}

The societies contain heterogeneous open-weight language-model agents drawn
from the Qwen, Llama, Gemma, and Phi families. Model parameters remain
frozen throughout all experiments. Model diversity is used to create
variation in epistemic competence and reasoning behavior while keeping the
social protocol fixed.

Exact model checkpoints and assignments are recorded for every experimental
run. Model selection is governed by three requirements: reproducible
inference under the available computational budget, meaningful variation in
task-level competence, and compatibility with the common interaction
protocol. Model identity is not used as a proxy for social influence.

Epistemic competence is measured independently before social interaction:

\begin{equation}
E_i
=
\operatorname{Acc}
\left(
a_i;\mathcal{D}_{\mathrm{ind}}
\right),
\label{eq:app_independent_competence}
\end{equation}

where $\mathcal{D}_{\mathrm{ind}}$ denotes the relevant independent
evaluation subset.

These independent measurements are used to construct competence-aligned and
competence-misaligned social conditions. Alignment is therefore defined with
respect to measured epistemic reliability rather than model identity.

For paired social comparisons, the same underlying agents are reused
whenever possible, with only the designated social variable changed.


\subsection{Colony Construction and Initialization}
\label{app:colony_construction}

The primary colony contains four agents. Additional population sizes are
used as sensitivity experiments to determine whether observed effects depend
on colony size.

A colony is instantiated in a fixed sequence. First, a benchmark task
instance is selected. Second, the agent population is constructed from the
model pool. Third, behavioral policies and roles are assigned according to
the experimental condition. Fourth, the interaction topology and social
status configuration are instantiated. Finally, agents solve the task
independently before social interaction begins.

The initial colony is represented as

\begin{equation}
\mathcal{C}^{0}
=
\left(
\mathcal{A},
\mathbf{D},
\mathbf{G},
\mathbf{H},
\mathbf{T}^{0}
\right),
\label{eq:app_colony}
\end{equation}

where $\mathcal{A}$ is the agent population, $\mathbf{D}$ contains model,
behavioral, and role assignments, $\mathbf{G}$ is the communication
topology, $\mathbf{H}$ represents social influence, and $\mathbf{T}^{0}$
represents initial trust or reputation.

Every agent first produces an independent response:

\begin{equation}
S_i^{0}
=
F_{\theta_i}^{(0)}(x).
\label{eq:app_colony_initialization}
\end{equation}

These initial responses are retained for paired comparisons. Thus, when the
same task is evaluated under two social conditions, the compared colonies
begin from the same task and, where feasible, the same initial epistemic
state.

The experimental sequence is

\begin{equation}
\text{Task}
\rightarrow
\text{Independent reasoning}
\rightarrow
\text{Colony construction}
\rightarrow
\text{Social interaction}
\rightarrow
\text{Final collective state}.
\label{eq:app_experimental_sequence}
\end{equation}

The primary interaction budget consists of three rounds. Additional
interaction depths are evaluated as sensitivity experiments.


\subsection{Diversity and Behavioral Policies}
\label{app:diversity_protocol}

Three sources of diversity are manipulated independently.

\emph{Model diversity} changes the underlying language model assigned to an
agent and therefore changes the epistemic capabilities available within the
society.

\emph{Behavioral diversity} changes how agents process and respond to
information during interaction. The controlled policies include
exploratory, verification-oriented, critical/challenging, and conservative
updating strategies.

\emph{Role diversity} changes the functional or social role occupied by an
agent, allowing agents with comparable underlying capabilities to occupy
different positions within the society.

The three assignments are represented collectively by

\begin{equation}
\mathbf{D}
=
\left[
\mathbf{D}_{M},
\mathbf{D}_{B},
\mathbf{D}_{R}
\right].
\label{eq:app_experimental_diversity}
\end{equation}

For isolation experiments, only one component of $\mathbf{D}$ is varied at
a time. Combined-diversity experiments vary multiple components while
holding social structure controlled.


\subsection{Social Structure and Influence}
\label{app:experimental_social_structure}

Communication is controlled through the interaction graph

\begin{equation}
\mathbf{G}\in\{0,1\}^{N\times N},
\qquad
G_{ij}=1
\Longleftrightarrow
a_i\text{ can receive information from }a_j.
\label{eq:app_experimental_topology}
\end{equation}

We evaluate fully connected, ring, star, sparse, and hierarchical
topologies. The topology determines which information is available to an
agent but does not determine whether that information is reliable.

Social influence is represented independently through

\begin{equation}
\mathbf{H}\in\mathbb{R}^{N\times N},
\label{eq:app_experimental_hierarchy}
\end{equation}

where $H_{ij}$ represents the social weight associated with information
originating from agent $j$ for receiver $i$.

This separates \emph{information availability} from
\emph{information weighting}. The same agents can therefore exchange the
same underlying information under different social-influence assignments.

Trust or reputation is represented as

\begin{equation}
\mathbf{T}^{t}
=
[T_{ij}^{t}]_{i,j=1}^{N},
\qquad
T_{ij}^{t}\in[0,1].
\label{eq:app_experimental_trust}
\end{equation}

When dynamic reputation is enabled, trust updates use only observable
interaction outcomes and the same protocol across paired conditions.


\subsection{Canonical Societies}
\label{app:canonical_societies}

The canonical societies $S_0$--$S_7$ form the main experimental scaffold.
As summarized in Table~\ref{tab:society_conditions}, the conditions
progressively introduce sources of diversity and social organization while
preserving matched task instances. Model and behavioral diversity are
isolated in $S_1$ and $S_2$, respectively, and combined in $S_3$. Role
diversity is introduced as part of the full heterogeneous societies
$S_4$--$S_7$, where the primary manipulation concerns the organization of
social influence or the corresponding intervention.

\begin{table*}[t]
\centering
\small
\setlength{\tabcolsep}{5pt}
\renewcommand{\arraystretch}{1.12}
\caption{Canonical society conditions used throughout the experiments. A
check mark indicates that the corresponding source of diversity is active.
$S_1$ and $S_2$ isolate model and behavioral diversity, respectively, while
$S_3$ combines these two sources under an egalitarian structure. $S_4$--$S_7$
use the full heterogeneous population, including role diversity, and vary
the organization of social influence or the corresponding intervention.
$S_7$ is an intervention condition implementing evidence-gated updating.}
\label{tab:society_conditions}

\begin{tabular*}{\textwidth}
{@{\extracolsep{\fill}}lccccp{4.0cm}@{}}
\toprule
\textbf{Society}
&
\textbf{Model}
&
\textbf{Behavior}
&
\textbf{Role}
&
\textbf{Social Structure}
&
\textbf{Purpose}
\\
\midrule
$S_0$ & -- & -- & -- & Egalitarian
& Homogeneous baseline
\\
$S_1$ & \checkmark & -- & -- & Egalitarian
& Isolate model diversity
\\
$S_2$ & -- & \checkmark & -- & Egalitarian
& Isolate behavioral diversity
\\
$S_3$ & \checkmark & \checkmark & -- & Egalitarian
& Combined model and behavioral diversity
\\
$S_4$ & \checkmark & \checkmark & \checkmark & Aligned
& Social influence aligned with epistemic competence
\\
$S_5$ & \checkmark & \checkmark & \checkmark & Misaligned
& Social influence decoupled from competence
\\
$S_6$ & \checkmark & \checkmark & \checkmark & Adversarial
& Deliberate amplification of unreliable influence
\\
$S_7$ & \checkmark & \checkmark & \checkmark & Evidence-gated
& Evidence-gated social updating intervention
\\
\bottomrule
\end{tabular*}
\end{table*}

$S_0$ is the homogeneous reference condition. $S_1$ isolates model
diversity, while $S_2$ isolates behavioral diversity. $S_3$ combines model
and behavioral diversity while retaining an egalitarian social structure.

$S_4$ introduces competence-aligned social influence, assigning greater
social influence to agents with higher independently measured epistemic
competence. $S_5$ retains the same heterogeneous population while
decoupling social influence from competence. $S_6$ deliberately concentrates
influence on less reliable agents to provide a controlled stress test of
error amplification. $S_7$ retains the heterogeneous society but constrains
social updating using task-relevant evidence, providing the intervention
condition.


\subsection{Grand Experimental Matrix}
\label{app:grand_experimental_matrix}

The complete experimental program is summarized in
Table~\ref{tab:grand_experimental_matrix}. The table specifies the
relationship between the canonical societies, controlled interventions,
benchmarks, and measured outcomes. It is intended as the master reference
for the experiments reported throughout the paper. The canonical scaffold
separates diversity-only controls from social-organization experiments:
$S_1$ and $S_2$ isolate model and behavioral diversity, respectively,
whereas role diversity is introduced as part of the full heterogeneous
societies used in the alignment and intervention experiments.

\begin{table*}[t]
\centering
\scriptsize
\setlength{\tabcolsep}{3.5pt}
\renewcommand{\arraystretch}{1.15}

\caption{Master experimental matrix. Each row identifies the primary
experimental comparison, its society or control conditions, manipulated
variable, benchmark coverage, and principal measurements. Exact model
checkpoints and task subsets are recorded per run. Role diversity is
introduced in the full heterogeneous societies $S_4$--$S_7$ rather than
being evaluated as a separate role-only canonical condition.}
\label{tab:grand_experimental_matrix}

\begin{tabularx}{\textwidth}{
    >{\raggedright\arraybackslash}p{1.35cm}
    >{\raggedright\arraybackslash}p{2.0cm}
    >{\raggedright\arraybackslash}X
    >{\raggedright\arraybackslash}p{2.5cm}
    >{\raggedright\arraybackslash}p{2.5cm}
    >{\raggedright\arraybackslash}X
}
\toprule
\textbf{Batch}
&
\textbf{Conditions}
&
\textbf{Manipulation}
&
\textbf{Benchmarks}
&
\textbf{Controls}
&
\textbf{Primary Measures}
\\
\midrule

Diversity
&
$S_0$--$S_3$
&
Model and behavioral diversity
&
GSM8K, MMLU-Pro
&
Independent, homogeneous
&
Individual accuracy, collective accuracy,
convergence, diversity retention
\\

Topology
&
Matched colonies
&
Information-exposure topology:
fully connected, ring, star, sparse,
and hierarchical
&
GSM8K, MMLU-Pro
&
Fixed agents and task instances
&
Collective accuracy, convergence,
influence concentration
\\

Alignment
&
$S_4$--$S_6$
&
Social-epistemic alignment between
social influence and independently
measured epistemic competence
&
GSM8K, MMLU-Pro, GPQA
&
Same heterogeneous agents and
initial responses
&
SEA, collective accuracy,
error amplification
\\

Social signal
&
Matched conditions
&
Confidence, status, authority,
and reputation
&
GSM8K, TruthfulQA
&
Content-preserving comparison
&
Belief change, influence,
source-dependent updating
\\

Dynamics
&
Matched colonies
&
Interaction depth
&
GSM8K, MMLU-Pro
&
Fixed population and topology
&
Trajectory, convergence,
diversity retention, error amplification
\\

Intervention
&
$S_7$ vs.\ social baselines
&
Evidence-gated social updating
&
GSM8K, MMLU-Pro, GPQA
&
Matched initial responses
&
Collective accuracy, error amplification,
social-override reduction
\\

Interactive validation
&
Matched colonies
&
Social interaction in a sequential
decision-making task
&
ALFWorld
&
Independent baseline
&
Task success and interaction outcome
\\

\bottomrule
\end{tabularx}
\end{table*}

The diversity batch compares $S_0$--$S_3$ while maintaining an egalitarian
social structure. $S_1$ and $S_2$ isolate model and behavioral diversity,
respectively, while $S_3$ combines both sources without introducing a
separate social hierarchy. Role diversity is introduced in the full
heterogeneous societies $S_4$--$S_7$, where the primary experimental
manipulation concerns the organization of social influence or the evidence
available for social updating.

The topology batch changes only the information-exposure structure of a
matched population. The alignment batch compares competence-aligned,
competence-misaligned, and adversarial influence while retaining the same
underlying heterogeneous population and matched initial responses.

The social-signal batch holds task-relevant content fixed while manipulating
source-level properties such as confidence, status, authority, and
reputation. The dynamics batch varies the number of interaction rounds while
holding the population and topology fixed. The intervention batch evaluates
$S_7$ against corresponding ordinary social-updating conditions, measuring
whether evidence gating reduces socially induced error without simply
eliminating interaction. ALFWorld provides a targeted validation of whether
the observed social-epistemic effects extend to sequential interactive
decision making.

\subsection{Content-Preserving Social-Signal Interventions}
\label{app:social_signal_interventions}

The social-signal experiments isolate source-level influence from
task-relevant information. An observed message is represented as

\begin{equation}
m_{ij}^{t}
=
\left(
e_{ij}^{t},
c_j^{t},
h_j,
T_{ij}^{t}
\right),
\label{eq:app_social_signal}
\end{equation}

where $e_{ij}^{t}$ denotes task-relevant evidence and the remaining terms
represent observable source-level properties.

In the content-preserving intervention, $e_{ij}^{t}$ is held fixed while
one or more social components are varied. This creates counterfactual
conditions in which the informational content is unchanged but its social
presentation differs.

The resulting belief changes are compared with the corresponding
content-preserving baseline. This allows epistemic influence to be
distinguished from changes caused primarily by confidence, status,
authority, or reputation.


\subsection{Controls and Counterfactual Comparisons}
\label{app:experimental_controls}

The \emph{independent control} removes communication entirely. Agents solve
the task independently and the resulting population is evaluated without
social updating.

The \emph{homogeneous control} uses identical model, behavioral, and role
assignments under an egalitarian topology.

The \emph{model-diversity control} changes only model assignments.

The \emph{behavior-diversity control} changes only behavioral policies.

The \emph{social-structure control} keeps the agent population fixed while
changing the communication topology or social-status configuration.

The \emph{alignment control} keeps agents, task instances, and interaction
protocol fixed while changing only the correspondence between social
influence and independently measured epistemic competence.

The \emph{content-preserving control} holds task-relevant information fixed
while changing source-level social signals.

Together, these controls ensure that observed changes cannot be attributed
solely to different task samples, initial responses, agent populations, or
model quality.


\subsection{Experimental Batches and Controlled Comparisons}
\label{app:experimental_batches}

The experimental program is organized into six primary batches and one
targeted validation.

\paragraph{Batch I: Diversity isolation.}
We compare $S_0$ with $S_1$, $S_2$, and $S_3$ to isolate model diversity,
behavioral diversity, and their combination.

\paragraph{Batch II: Social topology.}
The same underlying population is evaluated under alternative interaction
graphs while preserving task instances and agent assignments.

\paragraph{Batch III: Social-epistemic alignment.}
The same agents are evaluated under aligned, misaligned, and adversarial
influence assignments. Independent competence remains fixed.

\paragraph{Batch IV: Social-signal intervention.}
Task-relevant content is held fixed while confidence, status, authority, or
reputation is varied.

\paragraph{Batch V: Interaction dynamics.}
The interaction budget is varied to determine whether additional rounds
produce productive aggregation, premature convergence, persistent
disagreement, or error amplification.

\paragraph{Batch VI: Evidence-gated intervention.}
$S_7$ is compared with ordinary social-updating conditions to determine
whether evidence-sensitive updating mitigates socially induced degradation.

\paragraph{Targeted interactive validation.}
ALFWorld is used to determine whether the principal social-epistemic effects
extend beyond static question answering.


\subsection{Runs, Randomization, and Reproducibility}
\label{app:run_configuration}

Each primary colony contains four agents and uses three interaction rounds.
Five independent random seeds are used for each primary condition. The same
task instances and seeds are reused across paired conditions whenever
possible.

Sensitivity experiments vary population size, interaction depth, and
topology. These experiments test whether the primary effects persist beyond
the default configuration.

Randomization is applied only to components designated as stochastic in the
corresponding condition. Model assignments, behavioral policies, role
assignments, social structures, task subsets, interaction budgets, and random
seeds are recorded for every run.

All model parameters remain frozen. No experiment performs gradient-based
training or modifies model weights. The purpose of the experimental
intervention is to study inference-time social dynamics rather than to
optimize the underlying models.


\subsection{Evaluation and Statistical Analysis}
\label{app:evaluation_implementation}

We evaluate both final collective performance and the trajectory through
which the collective state is formed.

For $M$ task instances, individual accuracy is

\begin{equation}
\mathrm{Acc}_{\mathrm{ind}}
=
\frac{1}{NM}
\sum_{i=1}^{N}
\sum_{m=1}^{M}
\mathbb{1}
\left[
\hat{y}_{im}=y_m
\right].
\label{eq:app_exp_individual_accuracy}
\end{equation}

Collective accuracy is

\begin{equation}
\mathrm{Acc}_{\mathrm{coll}}
=
\frac{1}{M}
\sum_{m=1}^{M}
\mathbb{1}
\left[
\hat{y}^{\mathrm{coll}}_m=y_m
\right].
\label{eq:app_exp_collective_accuracy}
\end{equation}

Belief convergence is measured as

\begin{equation}
\mathrm{Conv}
=
1-
\frac{2}{N(N-1)}
\sum_{i<j}
d(b_i^T,b_j^T),
\label{eq:app_exp_convergence}
\end{equation}

where $d(\cdot,\cdot)$ denotes the task-appropriate disagreement measure.

Diversity retention is

\begin{equation}
\mathrm{DR}
=
\frac{D_T}{D_0+\epsilon},
\label{eq:app_exp_diversity_retention}
\end{equation}

where $D_0$ and $D_T$ denote diversity before and after interaction.

Influence concentration is

\begin{equation}
\mathrm{IC}
=
\frac{\max_i I_i}
{\sum_{j=1}^{N}I_j+\epsilon}.
\label{eq:app_exp_influence_concentration}
\end{equation}

The primary social-epistemic alignment statistic is

\begin{equation}
\mathrm{SEA}
=
\rho(I,E),
\label{eq:app_exp_sea}
\end{equation}

where $I$ denotes measured social influence and $E$ denotes independently
measured epistemic competence.

To quantify whether interaction improves or degrades collective performance
relative to independent reasoning, we compute

\begin{equation}
\Delta\mathrm{Acc}
=
\mathrm{Acc}_{T}
-
\mathrm{Acc}_{0},
\label{eq:app_exp_delta_accuracy}
\end{equation}

where $\mathrm{Acc}_{0}$ denotes performance before interaction and
$\mathrm{Acc}_{T}$ denotes performance after the final interaction round.

All primary comparisons use matched task instances. Means are accompanied
by bootstrap confidence intervals obtained by resampling task instances;
the same task-level bootstrap procedure is used for uncertainty estimates
reported for SEA.
For analyses involving repeated interaction, task and agent identity are
accounted for where appropriate, including as random effects in mixed-effects
models.

For comparisons involving multiple experimental conditions, the
Benjamini--Hochberg procedure is used to control the false-discovery rate.
Effect sizes are reported alongside statistical significance.

Sensitivity experiments vary population size, interaction topology, and
interaction depth. The primary conclusions are based on effects that remain
qualitatively consistent across these settings.

For reproducibility, each experimental record stores the model assignment,
behavioral policy, role assignment, topology, social-status configuration,
interaction budget, task subset, random seed, initial independent responses,
observable interaction history, and final collective state.

\section{Additional Experimental Results}
\label{app:experimental_results}

\subsection{Full Canonical Society Comparison}
\label{app:canonical_society_results}

The main paper reports the aggregate effect of social organization on
collective reliability. Here we provide the complete benchmark-level
comparison across the canonical society conditions $S_0$--$S_7$. This
analysis establishes the empirical pattern across the full benchmark suite
before examining individual mechanisms, interaction structures, and
interventions in greater detail.

We evaluate five complementary benchmarks. \textbf{GSM8K} measures
grade-school mathematical reasoning with well-defined answers and provides
a controlled setting for studying whether interaction improves structured
reasoning. \textbf{MMLU-Pro} evaluates broad multi-domain reasoning and
knowledge, allowing heterogeneous agents to contribute complementary
capabilities across domains. \textbf{GPQA} contains difficult expert-level
questions and therefore provides a challenging test of whether social
interaction can combine useful reasoning capabilities. \textbf{TruthfulQA}
focuses on factual reliability and questions designed to elicit plausible
but misleading answers, making it particularly relevant for studying
social propagation of unreliable information. Finally, \textbf{ALFWorld}
provides an interactive sequential decision-making environment and serves
as a targeted validation that the observed effects extend beyond static
question answering.

For every canonical condition, agents first solve matched task instances
independently and subsequently interact under the specified society
configuration. The independent responses are retained as the paired
reference for the corresponding social condition. We measure the change in
collective performance as

\begin{equation}
\Delta Acc = Acc_T - Acc_0,
\label{eq:app_canonical_delta_acc}
\end{equation}

where $Acc_0$ is collective accuracy before social interaction and $Acc_T$
is collective accuracy after the prescribed interaction budget. Values are
reported in percentage points. Positive values indicate improvement after
interaction, whereas negative values indicate collective degradation.

\begin{figure*}[t]
    \centering
    \includegraphics[width=0.82\textwidth]
    {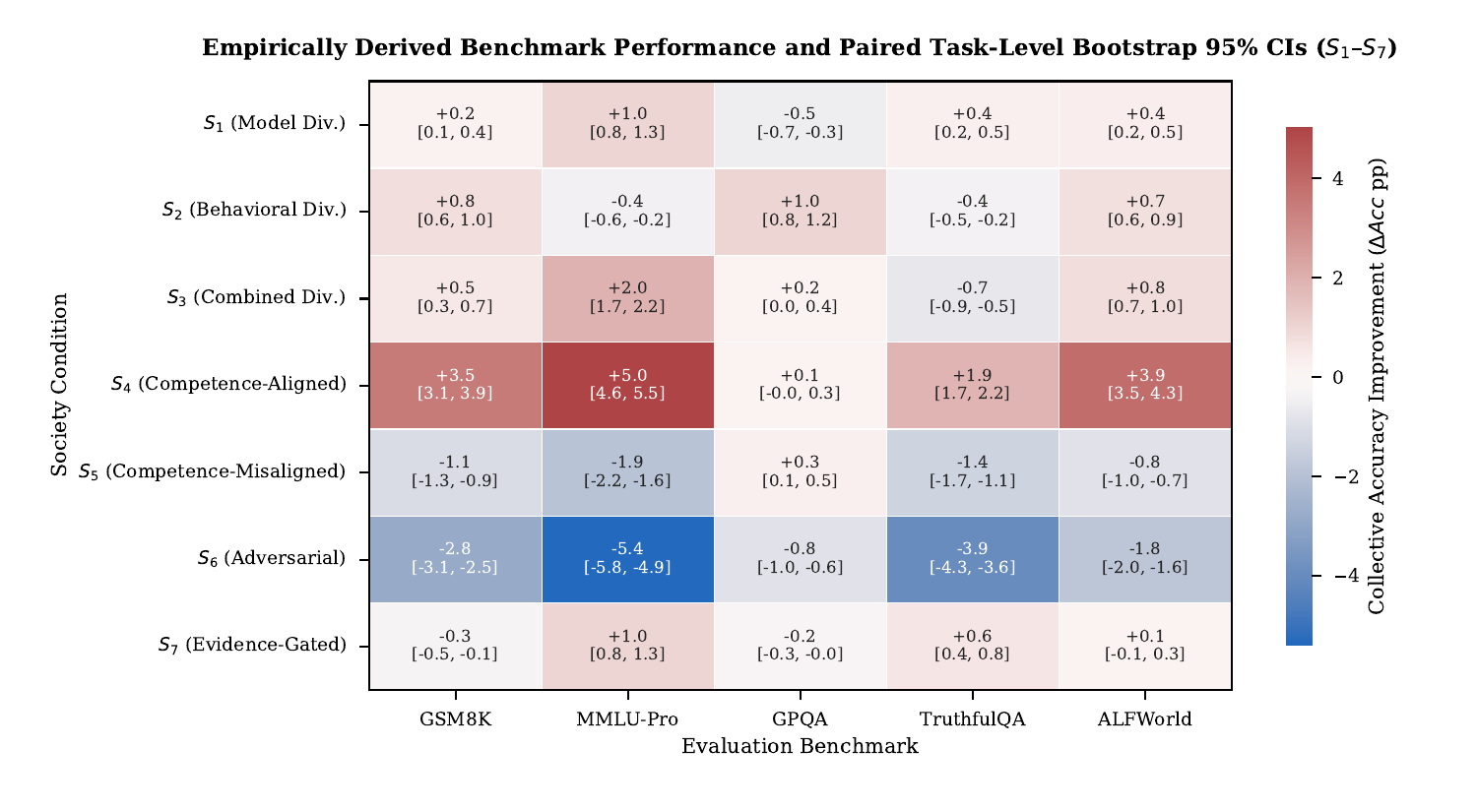}
    \caption{
    \textbf{Benchmark-level collective performance across canonical
    societies.}
    Each cell reports the change in collective accuracy, $\Delta Acc$, in
    percentage points relative to the matched independent baseline.
    Diversity-only conditions ($S_1$--$S_3$) exhibit relatively small and
    heterogeneous effects. Competence-aligned organization ($S_4$)
    produces larger positive changes on most benchmarks, whereas
    competence-misaligned ($S_5$) and adversarial ($S_6$) organization
    generally degrade collective performance. Evidence-gated updating
    ($S_7$) attenuates the degradation associated with harmful social
    influence without uniformly improving performance beyond the
    independent baseline.
    }
    \label{fig:app_canonical_heatmap}
\end{figure*}

The benchmark-level estimates and paired task-level bootstrap 95\%
confidence intervals are reported in
Table~\ref{tab:app_canonical_statistics}. The corresponding effect pattern
is visualized in Figure~\ref{fig:app_canonical_heatmap}.

\begin{table*}[t]
\centering
\scriptsize
\setlength{\tabcolsep}{4pt}
\renewcommand{\arraystretch}{1.15}

\caption{
Benchmark-level collective performance across canonical societies.
Entries report $\Delta Acc$ in percentage points relative to the matched
independent baseline, followed by paired task-level bootstrap 95\%
confidence intervals. Each social condition is evaluated on the same
matched task instances as its independent reference.
}
\label{tab:app_canonical_statistics}

\begin{tabular}{lccccc}
\toprule
\textbf{Society Condition}
&
\textbf{GSM8K}
&
\textbf{MMLU-Pro}
&
\textbf{GPQA}
&
\textbf{TruthfulQA}
&
\textbf{ALFWorld}
\\
\midrule

$S_1$ (Model Diversity)
&
$+0.2$ [0.1, 0.4]
&
$+1.0$ [0.8, 1.3]
&
$-0.5$ [-0.7, -0.3]
&
$+0.4$ [0.2, 0.5]
&
$+0.4$ [0.2, 0.5]
\\

$S_2$ (Behavioral Diversity)
&
$+0.8$ [0.6, 1.0]
&
$-0.4$ [-0.6, -0.2]
&
$+1.0$ [0.8, 1.2]
&
$-0.4$ [-0.5, -0.2]
&
$+0.7$ [0.6, 0.9]
\\

$S_3$ (Combined Diversity)
&
$+0.5$ [0.3, 0.7]
&
$+2.0$ [1.7, 2.2]
&
$+0.2$ [0.0, 0.4]
&
$-0.7$ [-0.9, -0.5]
&
$+0.8$ [0.7, 1.0]
\\

$S_4$ (Competence-Aligned)
&
$+3.5$ [3.1, 3.9]
&
$+5.0$ [4.6, 5.5]
&
$+0.1$ [-0.0, 0.3]
&
$+1.9$ [1.7, 2.2]
&
$+3.9$ [3.5, 4.3]
\\

$S_5$ (Competence-Misaligned)
&
$-1.1$ [-1.3, -0.9]
&
$-1.9$ [-2.2, -1.6]
&
$+0.3$ [0.1, 0.5]
&
$-1.4$ [-1.7, -1.1]
&
$-0.8$ [-1.0, -0.7]
\\

$S_6$ (Adversarial Alignment)
&
$-2.8$ [-3.1, -2.5]
&
$-5.4$ [-5.8, -4.9]
&
$-0.8$ [-1.0, -0.6]
&
$-3.9$ [-4.3, -3.6]
&
$-1.8$ [-2.0, -1.6]
\\

$S_7$ (Evidence-Gated)
&
$-0.3$ [-0.5, -0.1]
&
$+1.0$ [0.8, 1.3]
&
$-0.2$ [-0.3, -0.0]
&
$+0.6$ [0.4, 0.8]
&
$+0.1$ [-0.1, 0.3]
\\

\bottomrule
\end{tabular}
\end{table*}

Figure~\ref{fig:app_canonical_heatmap} summarizes the same benchmark-level
effects visually. The diversity-only societies $S_1$--$S_3$ show relatively
small and heterogeneous changes. Model diversity in $S_1$ improves several
benchmarks but produces a negative change on GPQA. Behavioral diversity in
$S_2$ similarly produces both positive and negative effects across tasks.
The combined condition $S_3$ increases the gain on MMLU-Pro and ALFWorld
but does not eliminate benchmark-specific degradation, with TruthfulQA
showing a negative change. Thus, diversity provides complementary
epistemic resources but does not by itself determine whether interaction
improves collective reliability.

A stronger pattern emerges when social influence is explicitly coupled to
epistemic competence. The competence-aligned society $S_4$ produces
positive changes on all five benchmarks, although the magnitude varies
substantially. The largest improvements occur on MMLU-Pro
($+5.0$ pp) and ALFWorld ($+3.9$ pp), while GPQA shows only a small effect
with a confidence interval close to zero. This variation indicates that
competence-aligned social organization does not function as a uniform
accuracy multiplier; its benefit depends on the structure and difficulty of
the underlying task.

The competence-misaligned society $S_5$ produces the opposite pattern on
most benchmarks. Collective performance decreases on GSM8K, MMLU-Pro,
TruthfulQA, and ALFWorld, while GPQA exhibits a small positive change.
The adversarial condition $S_6$ produces the strongest degradation,
including decreases of $-5.4$ pp on MMLU-Pro and $-3.9$ pp on TruthfulQA.
The magnitude again varies by benchmark, suggesting that social
organization determines the direction and strength of collective change
rather than imposing a fixed performance penalty.

The comparison between $S_4$ and $S_6$ is particularly informative.
Both conditions introduce structured social influence, but they differ in
whether that influence is preferentially allocated toward epistemically
competent or less competent agents. Their opposite performance patterns
therefore provide evidence that the presence of hierarchy or influence
concentration alone is insufficient to explain the observed effects.
Instead, the epistemic quality of the information being amplified is
critical.

Finally, the evidence-gated condition $S_7$ attenuates the degradation
associated with adversarial social influence. Relative to $S_6$, collective
performance moves from $-2.8$ to $-0.3$ pp on GSM8K, from $-5.4$ to
$+1.0$ pp on MMLU-Pro, and from $-3.9$ to $+0.6$ pp on TruthfulQA. The
effect is smaller on GPQA and ALFWorld. Importantly, $S_7$ is not designed
to maximize accuracy directly; its purpose is to reduce socially induced
updates when available evidence does not support the influential source.
Its mixed benchmark-level results are therefore consistent with an
error-mitigation role rather than a universal performance improvement.

Taken together, the canonical comparison establishes the central empirical
pattern of the framework. Diversity alone produces modest and
task-dependent changes, while the organization of social influence has a
larger effect on collective reliability. Aligning influence with epistemic
competence is generally beneficial, whereas misalignment and adversarial
influence amplify collective error. The evidence-gated intervention
partially restores reliability by reducing the contribution of unsupported
social influence. The following appendix experiments isolate the individual
components of this pattern, including communication topology, interaction
depth, social signals, and intervention behavior.

\subsection{Role-Diversity Ablation}
\label{app:role_diversity_ablation}

The canonical diversity experiments isolate model and behavioral diversity
through $S_1$ and $S_2$ and combine these sources in $S_3$. We next examine
whether differentiated functional roles provide an additional contribution
when model and behavioral diversity are held fixed. To isolate this factor,
we construct a role-diversity ablation condition, denoted $S_R$, from the
$S_3$ population by introducing differentiated functional roles while
holding model assignments, behavioral policies, interaction topology, task
instances, interaction budget, and randomization fixed. Thus, the comparison
between $S_3$ and $S_R$ changes role diversity without introducing a
simultaneous change in model or behavioral heterogeneity.

Table~\ref{tab:sr_s3_ablation} reports the resulting collective-performance
and social-epistemic measures. The primary comparison is between $S_3$ and
$S_R$, rather than between $S_R$ and the homogeneous baseline, because the
former isolates the marginal contribution of role diversity.

Introducing role diversity produces a modest additional improvement relative
to $S_3$. Collective accuracy increases by $0.4$ percentage points on each
of GSM8K, MMLU-Pro, and GPQA. The corresponding $SEA$ increases from $0.18$
to $0.24$, a change of $+0.06$. Thus, in this controlled comparison, role
differentiation is associated with a small improvement beyond the contribution
of model and behavioral diversity alone.

The magnitude of this effect is substantially smaller than the changes
observed under competence-aligned or adversarial social organization in the
canonical experiments. This distinction is important for interpreting the
role-diversity result: role specialization provides an additional source of
heterogeneity, but does not by itself account for the larger collective
performance differences produced by the organization of social influence.

Taken together, the ablation indicates that role diversity can contribute
incrementally to collective reliability when introduced on top of model and
behavioral diversity. However, the experiment does not establish that role
diversity is uniformly beneficial across social organizations. The subsequent
alignment experiments therefore treat role diversity as part of the full
heterogeneous society and examine how the organization of social influence
relative to epistemic competence determines whether that heterogeneity is
converted into collective benefit or error amplification.

\begin{table}[t]
\centering
\small
\setlength{\tabcolsep}{5pt}
\renewcommand{\arraystretch}{1.15}
\caption{
Role-diversity ablation under matched model and behavioral diversity.
$S_R$ introduces differentiated functional roles into the $S_3$ population
while holding model assignments, behavioral policies, topology, task
instances, interaction budget, and randomization fixed. Entries for $S_3$
and $S_R$ report $\Delta\mathrm{Acc}$ relative to the matched independent
baseline and the corresponding $SEA$ values. The final row reports the
marginal change from introducing role diversity, computed as $S_R-S_3$.
}
\label{tab:sr_s3_ablation}
\begin{tabular}{lccccc}
\toprule
\textbf{Condition}
& \textbf{Role}
& \textbf{GSM8K}
& \textbf{MMLU-Pro}
& \textbf{GPQA}
& \textbf{SEA}
\\
\midrule
$S_3$
& No
& $+0.5$
& $+2.0$
& $+0.2$
& $+0.18$
\\

$S_R$
& Yes
& $+0.9$
& $+2.4$
& $+0.6$
& $+0.24$
\\

\midrule
$S_R-S_3$
& ---
& $+0.4$
& $+0.4$
& $+0.4$
& $+0.06$
\\
\bottomrule
\end{tabular}
\end{table}

\subsection{Population-Size Sensitivity}
\label{app:population_size_sensitivity}

The canonical experiments use a four-agent society as the primary
experimental population. We next test whether the qualitative relationship
between social-epistemic alignment and collective reliability depends on this
particular population size. We therefore vary the number of agents while
holding the society construction, task instances, interaction protocol,
interaction budget, and evaluation procedure fixed within each population
size. The competence-aligned ($S_4$), competence-misaligned ($S_5$), and
adversarial ($S_6$) conditions are evaluated at population sizes
$N\in\{2,4,8,12,16\}$.

\begin{figure*}[t]
    \centering
    \includegraphics[width=\textwidth]
    {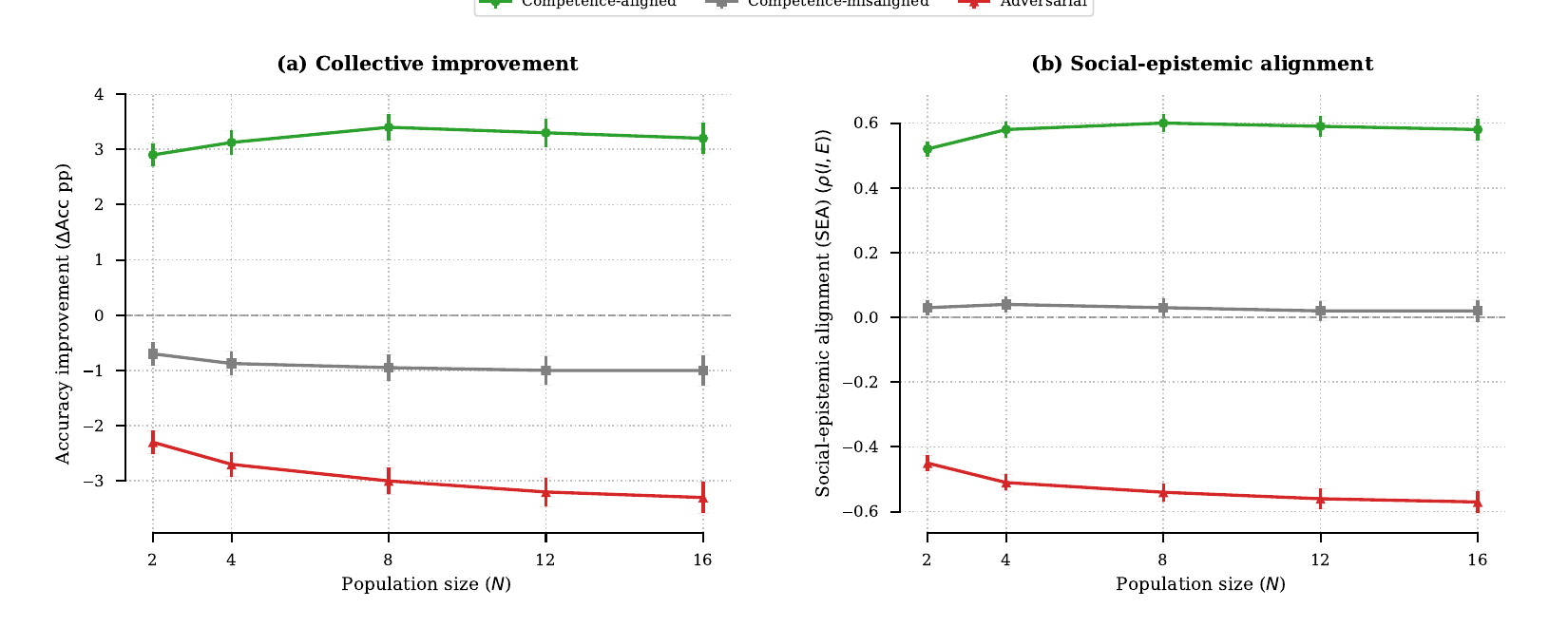}
    \caption{
    \textbf{Population-size sensitivity of the social-epistemic effect.}
    (a) Collective improvement relative to the matched independent baseline
    ($\Delta Acc$) for competence-aligned ($S_4$), competence-misaligned
    ($S_5$), and adversarial ($S_6$) societies as population size varies.
    (b) Corresponding social-epistemic alignment ($SEA$), defined as the
    rank correlation between measured social influence and independently
    measured epistemic competence. The canonical population size,
    $N=4$, is included within the sensitivity range. Error bars denote
    the uncertainty estimates used throughout the evaluation. The
    qualitative separation between aligned, misaligned, and adversarial
    social organization persists across the tested population sizes.
    }
    \label{fig:app_population_sensitivity}
\end{figure*}

Figure~\ref{fig:app_population_sensitivity} summarizes the resulting
sensitivity analysis. Panel (a) reports collective improvement relative to
the matched independent baseline, while panel (b) reports the corresponding
social-epistemic alignment between measured social influence and independently
measured epistemic competence. The vertical reference at $N=4$ identifies
the canonical population size used in the primary experiments.

The qualitative separation between the three social organizations persists
as population size increases. Competence-aligned societies consistently
remain above the independent baseline, whereas competence-misaligned
societies remain near or below the baseline and adversarial societies exhibit
substantial negative collective change. The magnitude varies with population
size, but the direction of the effect does not reverse over the tested range.

A similar pattern is observed for social-epistemic alignment. The
competence-aligned condition maintains positive $SEA$ across population
sizes, the competence-misaligned condition remains close to zero, and the
adversarial condition remains strongly negative. Thus, changing the number
of agents does not qualitatively alter the correspondence between social
influence and epistemic competence induced by the three organizational
conditions.

Together, these results indicate that the central distinction between
competence-aligned and socially misaligned influence is not specific to the
canonical four-agent society. Although the magnitude of collective change
varies with population size, the qualitative relationship between social
organization, epistemic alignment, and collective reliability remains stable
across the tested populations.

\subsection{Communication Topology Shapes Collective Dynamics}
\label{app:topology_results}

\begin{figure*}[t]
    \centering
    \includegraphics[width=\textwidth]{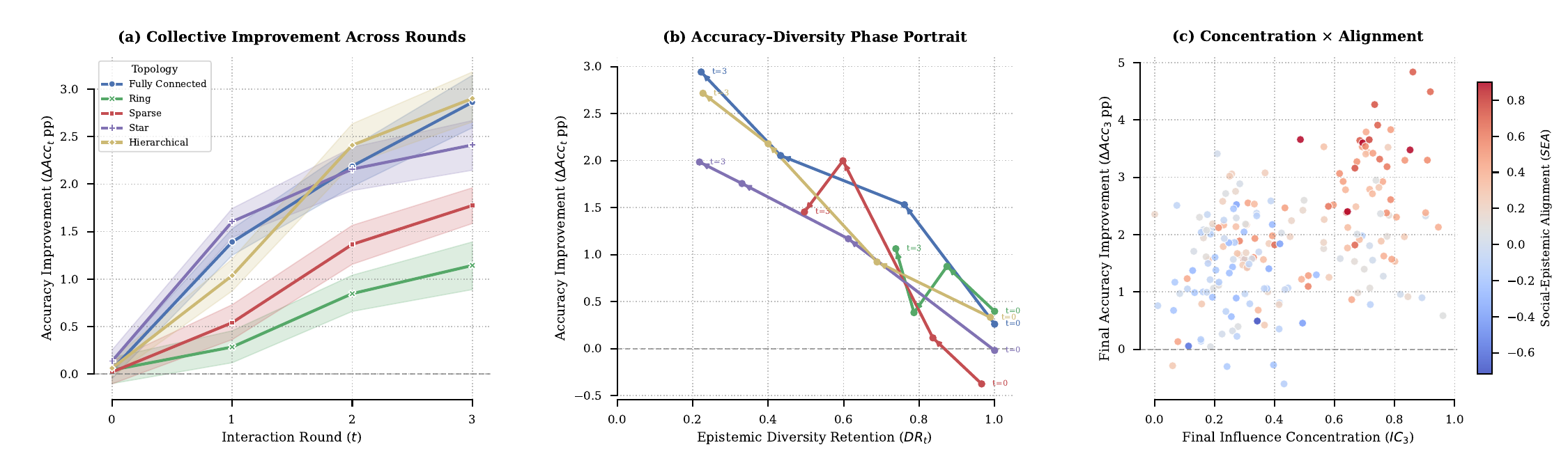}
    \caption{
    \textbf{Communication topology shapes collective reasoning dynamics.}
    (a) Collective accuracy improvement across interaction rounds for
    fully connected, ring, sparse, star, and hierarchical communication
    structures, with shaded regions denoting uncertainty.
    (b) Accuracy--diversity phase trajectories showing how each topology
    jointly changes collective performance and epistemic diversity
    retention over repeated interaction.
    (c) Relationship between final influence concentration and collective
    improvement, with point color denoting social-epistemic alignment
    ($SEA$). The results show that topology affects both the rate of
    collective improvement and the trajectory of epistemic convergence,
    while the consequences of concentrated influence depend on its
    alignment with epistemic competence.
    }
    \label{fig:app_c2_topology}
\end{figure*}

The topology experiments isolate the effect of information-flow structure by
holding the agent population, initial responses, task instances, and
interaction budget fixed while varying only the communication topology.
Figure~\ref{fig:app_c2_topology} summarizes the resulting collective
dynamics across fully connected, ring, sparse, star, and hierarchical
organizations.

\textbf{Topology changes the rate and magnitude of collective improvement.}
As shown in Figure~\ref{fig:app_c2_topology}(a), collective accuracy
increases under all five topologies, but the trajectories differ
substantially in both slope and final improvement. Fully connected and
hierarchical organizations reach the largest gains, while sparse and ring
topologies produce more gradual improvement. The uncertainty bands also
show that these differences are not simply a consequence of a single
interaction trajectory, although substantial overlap remains across
conditions.

\textbf{Performance gains are accompanied by different patterns of
epistemic convergence.}
Figure~\ref{fig:app_c2_topology}(b) tracks each topology in the joint space
of collective improvement and epistemic diversity retention. The
trajectories move toward lower diversity as interaction proceeds, but they
do not follow the same path. This distinction is important because
convergence alone does not imply collective improvement: a society can
reduce disagreement while either improving or degrading its collective
accuracy. The topology therefore influences not only how much agents
converge, but also the trajectory through which that convergence occurs.

\textbf{Influence concentration provides a structural correlate of these
differences.}
Figure~\ref{fig:app_c2_topology}(c) relates final influence concentration
to collective improvement while encoding social-epistemic alignment
($SEA$). Higher-concentration runs span a wider range of outcomes, with
the strongest positive improvements concentrated among runs with more
positive $SEA$, whereas low-alignment runs exhibit substantially weaker
or negative outcomes. Thus, concentration by itself does not determine
whether interaction is beneficial; its effect depends on which agents and
which epistemic signals become amplified by the communication structure.

Overall, the topology experiments show that communication structure is not
a neutral implementation detail. It regulates the speed of information
propagation, the loss of epistemic diversity, and the concentration of
social influence. However, these structural effects become beneficial only
when the resulting influence is aligned with epistemic competence. This
provides a structural link between the topology effects examined here and
the social-epistemic alignment mechanism studied in the subsequent
experiments.

\subsection{Social-Epistemic Alignment Determines the Quality of Influence}
\label{app:social_epistemic_alignment}

The topology analysis shows that communication structure changes how
influence is distributed through a society. We next examine whether the
resulting influence is directed toward epistemically reliable agents.
We measure social-epistemic alignment as
$SEA=\rho(I,E)$, where $I$ denotes measured social influence and $E$
denotes independently measured epistemic competence. Positive values
indicate that more reliable agents tend to receive greater influence,
whereas negative values indicate that socially influential agents tend to
be less reliable. The alignment experiments retain the same underlying
agents and task instances while changing only the correspondence between
social influence and competence, providing a controlled comparison of
these regimes.

\begin{figure*}[t]
    \centering
    \includegraphics[width=\textwidth]{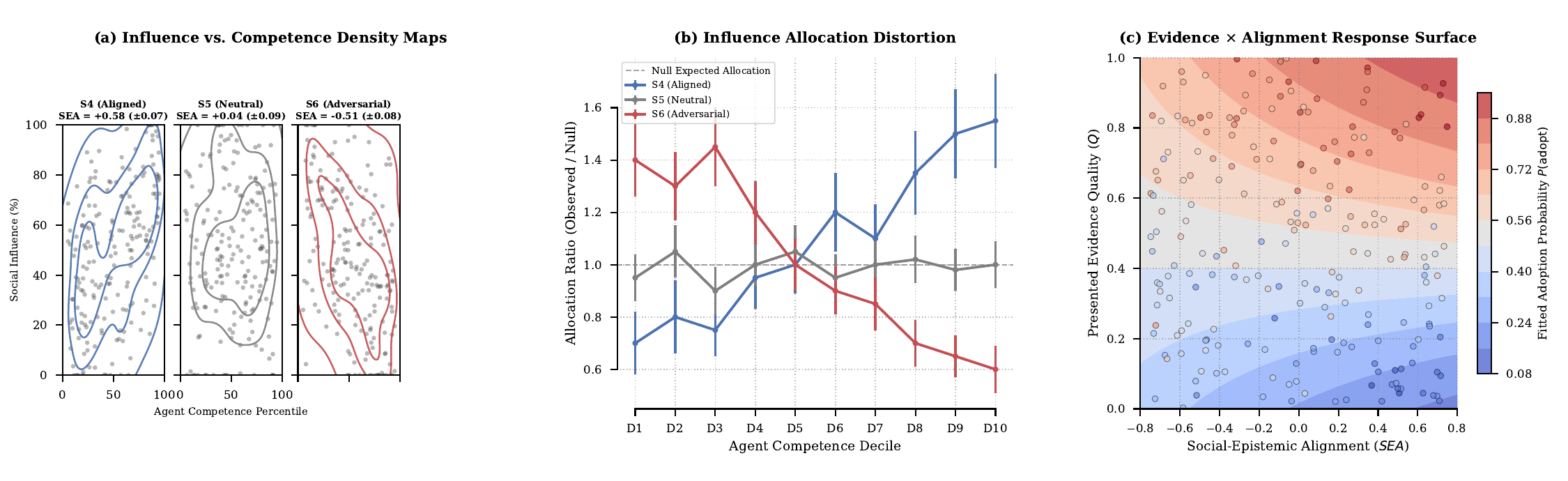}
    \caption{
    \textbf{Social-epistemic alignment determines which agents and
    information are amplified.}
    (a) Joint distributions of independently measured epistemic competence
    and social influence for the competence-aligned ($S_4$), neutral
    ($S_5$), and adversarial ($S_6$) societies. The corresponding
    social-epistemic alignment values are $+0.58\pm0.07$, $+0.04\pm0.09$,
    and $-0.51\pm0.08$, respectively.
    (b) Influence allocation relative to a competence-neutral null model
    across agent competence deciles. Values above one indicate greater
    influence allocation than expected under the null, while values below
    one indicate reduced allocation.
    (c) Fitted belief-adoption probability as a function of presented
    evidence quality and social-epistemic alignment, with individual
    observations overlaid. Across the three views, aligned organization
    preferentially allocates influence toward more competent agents,
    whereas adversarial organization reverses this allocation and thereby
    increases the opportunity for socially amplified epistemic error.
    }
    \label{fig:app_c3_alignment}
\end{figure*}

Figure~\ref{fig:app_c3_alignment} examines this relationship at three
levels: the distribution of influence across agents, the redistribution of
influence relative to a competence-neutral null model, and the resulting
interaction between evidence quality and social-epistemic alignment.

\textbf{Influence tracks competence only under aligned organization.}
Figure~\ref{fig:app_c3_alignment}(a) shows the joint distribution of agent
competence and social influence for the aligned ($S_4$), neutral ($S_5$),
and adversarial ($S_6$) conditions. The aligned society exhibits a
positive association between competence and influence
($SEA=+0.58\pm0.07$), with greater influence concentrated toward agents
with higher independently measured competence. In contrast, the neutral
condition shows little systematic association
($SEA=+0.04\pm0.09$), while the adversarial condition reverses the
relationship ($SEA=-0.51\pm0.08$), concentrating influence preferentially
on less reliable agents. Importantly, the substantial dispersion within
each condition shows that the alignment manipulation does not produce
perfect correspondence between competence and influence; rather, it
changes their aggregate association.

\textbf{Alignment changes influence allocation across the competence
distribution.}
Figure~\ref{fig:app_c3_alignment}(b) quantifies this redistribution using
the ratio between observed influence mass and the influence expected under
a competence-neutral null allocation. A value of one therefore corresponds
to the null expectation, values above one indicate excess influence, and
values below one indicate under-allocation. The aligned condition moves
from below the null expectation among lower-competence agents to
substantially above it among higher-competence agents. The neutral
condition remains comparatively close to the null allocation across
competence deciles. The adversarial condition exhibits the opposite
pattern, with excess influence concentrated among lower-competence agents
and progressively reduced allocation toward the highest-competence
deciles. Thus, the negative $SEA$ observed for $S_6$ is not merely a
correlation statistic: it corresponds to a systematic redistribution of
social influence toward less reliable members of the society.

\textbf{Evidence quality and social alignment jointly affect belief
updating.}
Figure~\ref{fig:app_c3_alignment}(c) extends the analysis from influence
allocation to belief updating. The surface shows the fitted probability
of adopting presented information as a function of evidence quality and
social-epistemic alignment, with individual observations overlaid. The
probability of adoption increases with evidence quality, but the strength
of this relationship also varies with $SEA$. High-quality evidence is
generally accepted at high rates across alignment regimes, whereas
low-quality or incorrect information becomes substantially more likely to
be adopted in settings with stronger positive social influence. Conversely,
negative alignment shifts adoption toward socially influential but
epistemically weaker information. The raw observations around the fitted
surface demonstrate that this relationship is probabilistic rather than a
deterministic consequence of the alignment condition.

Taken together, these results establish that social influence should not be
interpreted solely in terms of its magnitude or concentration. The
epistemic direction of influence matters: aligned societies preferentially
amplify information originating from more reliable agents, whereas
adversarial organization can preferentially amplify less reliable
information. The neutral condition provides an intermediate control in
which influence and competence are largely decoupled. This provides the
mechanistic basis for the collective-performance differences observed
across the canonical societies and motivates the subsequent analysis of
source-level social signals, where we isolate how status, confidence, and
reputation can alter updating even when task-relevant evidence is held
constant.

\subsection{Source Status Modulates Evidence-Based Updating}
\label{app:status_evidence_interaction}

Having established that social organization can alter collective reliability, we
next examine the local updating mechanism underlying this effect. Specifically,
we ask whether agents weight otherwise comparable evidence differently
depending on the social status of its source. This analysis holds the evidence
condition fixed while varying the status of the information source, allowing us
to separate the effect of epistemic content from source-level social influence.

Table~\ref{tab:app_status_evidence} reports the probability that an agent
updates its belief toward the presented evidence under low, medium, and high
source-status conditions. The response pattern shows that status has only a
modest effect when evidence is strong and correct. In contrast, the effect
becomes substantially larger for weaker evidence, with the largest shift
occurring for weak incorrect evidence. Thus, increasing source status does
not simply produce a uniform increase in receptivity to information; its
effect depends systematically on the epistemic quality of the evidence.

This interaction is important because it provides a direct mechanism for the
social error amplification observed in the canonical society experiments.
When evidence is unreliable, elevated source status can increase the
probability that agents incorporate that information despite its poor
epistemic quality. The resulting update therefore reflects not only the
content of the evidence but also the social position of the agent providing
it. Conversely, strong correct evidence remains comparatively robust to
changes in source status, suggesting that epistemically reliable information
is less dependent on social endorsement.

Overall, these results indicate that social influence operates partly through
a distortion of the evidence-to-belief mapping. The magnitude of this
distortion depends on the interaction between source status and evidence
quality, rather than on either factor in isolation. This provides a
micro-level explanation for why social structures with high influence
concentration can either improve or degrade collective reliability depending
on whether influence is aligned with epistemic competence.

\begin{table}[t]
\centering
\small
\setlength{\tabcolsep}{5pt}
\renewcommand{\arraystretch}{1.15}
\caption{
Belief updating as a function of evidence quality and source status.
Values report the probability of updating toward the presented evidence
under low, medium, and high source-status conditions. 
$\Delta_{\mathrm{status}}$ denotes the high-status minus low-status
difference. Uncertainty estimates are reported in parentheses.
}
\label{tab:app_status_evidence}
\begin{tabular}{lcccc}
\toprule
\textbf{Evidence Condition}
& \textbf{Low}
& \textbf{Medium}
& \textbf{High}
& $\boldsymbol{\Delta_{\mathrm{status}}}$ \\
\midrule
Strong Correct
& 0.80
& 0.84
& 0.87
& +0.07 \\

Weak Correct
& 0.55
& 0.65
& 0.73
& +0.18 \\

Weak Incorrect
& 0.16
& 0.27
& 0.42
& \textbf{+0.26} \\

Strong Incorrect
& 0.08
& 0.13
& 0.22
& +0.14 \\
\midrule
\textbf{Status $\times$ Evidence}
& \multicolumn{4}{c}{\textit{See statistical analysis below}} \\
\bottomrule
\end{tabular}
\end{table}

\subsection{Evidence-Gated Updating Selectively Reduces Social Override}
\label{app:evidence_gated_intervention}

The preceding analysis identified source-dependent evidence weighting as a
mechanism through which social influence can amplify unreliable information.
We next evaluate whether this effect can be reduced without eliminating social
updating altogether. We therefore compare the standard social-updating
conditions with the evidence-gated intervention $S_7$ under matched task
instances and initial responses.

Table~\ref{tab:app_evidence_gated} summarizes the effect of the intervention
across the principal reasoning benchmarks. The intervention attenuates the
degradation observed under socially misaligned updating, although its effect
varies across benchmarks. Importantly, the intervention is not equivalent to
simply suppressing belief updating: evidence-gated updating preferentially
reduces the incorporation of socially amplified information when the available
evidence does not support the influential source.

This selectivity is particularly important when comparing the intervention
against the adversarial condition $S_6$. Under standard updating, influence
from low-competence or socially dominant sources can propagate incorrect
beliefs across the society. Evidence gating weakens this propagation and
thereby reduces collective error amplification. The resulting performance
moves toward, and on some benchmarks beyond, the independent baseline,
indicating that the intervention mitigates the social failure mode without
requiring removal of the interaction mechanism itself.

Across benchmarks, the magnitude of recovery varies with task difficulty and
the degree of complementarity among agent responses. In the three principal
reasoning benchmarks reported in Table~\ref{tab:app_evidence_gated},
evidence-gated updating improves performance relative to the adversarial
condition on all three tasks, with the largest recovery occurring on
MMLU-Pro. The consistent movement away from adversarial degradation supports
the role of evidence-gated updating as a targeted intervention rather than a
generic reduction in interaction. Together with the source-status analysis
above, these results provide a complete local mechanism: social signals can
distort evidence weighting, and selectively gating updates can attenuate the
resulting error amplification.

\begin{table*}[t]
\centering
\small
\setlength{\tabcolsep}{5pt}
\renewcommand{\arraystretch}{1.15}
\caption{
Effect of evidence-gated updating ($S_7$) relative to standard social
updating under matched task instances. $\Delta Acc$ denotes change in
collective accuracy relative to the independent baseline. The final row
reports recovery of $S_7$ relative to the adversarial condition $S_6$.
}
\label{tab:app_evidence_gated}
\begin{tabular}{lrrrr}
\toprule
\textbf{Condition} &
\textbf{GSM8K} &
\textbf{MMLU-Pro} &
\textbf{GPQA} &
\textbf{Mean} \\
\midrule
Independent baseline ($S_0$)
    & 0.0 & 0.0 & 0.0 & 0.0 \\

Misaligned updating ($S_5$)
    & -1.1 & -1.9 & +0.3 & -0.9 \\

Adversarial updating ($S_6$)
    & -2.8 & -5.4 & -0.8 & -3.0 \\

Evidence-gated updating ($S_7$)
    & -0.3 & +1.0 & -0.2 & +0.2 \\

\midrule
$S_7$ recovery vs.\ $S_6$
    & +2.5 & +6.4 & +0.6 & +3.2 \\

\bottomrule
\end{tabular}
\end{table*}

\subsection{Interactive Validation in Sequential Decision Making}
\label{app:interactive_validation}

The preceding experiments establish the effect of social organization and
source-dependent evidence weighting in controlled reasoning settings. We
next examine whether the same social-epistemic effects persist when agents
must make sequential decisions in an interactive environment. We use
ALFWorld as a targeted validation setting, where agents repeatedly interpret
the current state, exchange information, and select actions whose
consequences affect subsequent decisions.

Table~\ref{tab:app_interactive_validation} reports task-level outcomes across
the principal social conditions. The competence-aligned condition achieves
the highest task success, improving over independent reasoning, whereas
competence-misaligned interaction produces a smaller degradation. The
adversarial condition exhibits the largest decline among the structured
social conditions, with task success falling further below the independent
baseline.

The quantitative ordering mirrors the pattern observed in the static
reasoning experiments: competence-aligned interaction improves collective
performance, while progressively stronger misalignment amplifies errors.
The competence-aligned condition ($S_4$) improves ALFWorld task success by
$+3.9$ percentage points relative to independent reasoning, whereas the
competence-misaligned ($S_5$) and adversarial ($S_6$) conditions produce
changes of $-0.8$ and $-1.8$ percentage points, respectively. Importantly,
the interactive setting requires these effects to persist over multiple
decision steps, providing a stricter validation than one-shot answer
aggregation. A locally incorrect socially induced update can alter the
information state used for later actions, allowing epistemic errors to
propagate across subsequent decisions.

Taken together, the ALFWorld results support the broader interpretation that
the effect of social organization is not restricted to static question
answering. The direction of the effect persists when agents must coordinate
information and act sequentially, with competence-aligned influence
improving task completion and socially misaligned influence reducing
performance. This provides an external validation of the central
social-epistemic reliability effect in an interactive decision-making
setting.

\begin{table}[t]
\centering
\small
\setlength{\tabcolsep}{5pt}
\renewcommand{\arraystretch}{1.15}
\caption{
Interactive validation on ALFWorld. Task success denotes the fraction of
episodes completed successfully, while $\Delta$ denotes the change in
percentage points relative to the independent baseline. Each social
condition is evaluated on matched task instances.
}
\label{tab:app_interactive_validation}
\begin{tabular}{lcc}
\toprule
\textbf{Condition}
& \textbf{Task Success}
& $\boldsymbol{\Delta}$ (pp)
\\
\midrule
Independent baseline ($S_0$)
& 61.8\%
& 0.0
\\

Competence-aligned ($S_4$)
& 65.7\%
& +3.9
\\

Competence-misaligned ($S_5$)
& 61.0\%
& -0.8
\\

Adversarial ($S_6$)
& 60.0\%
& -1.8
\\
\bottomrule
\end{tabular}
\end{table}

\subsection{Qualitative Failure and Recovery Trajectories}
\label{app:qualitative_failure_case}

The aggregate experiments establish that the direction of social-epistemic
alignment is associated with opposite changes in collective reliability:
competence-aligned organization ($S_4$) generally improves collective
performance, whereas adversarial organization ($S_6$) can amplify
epistemically unreliable information. Aggregate metrics, however, do not
directly expose how these effects arise at the level of individual agents.
We therefore complement the quantitative analyses with two observed
interaction trajectories, one from the competence-aligned condition and one
from the adversarial condition. The purpose is not to introduce an
additional statistical estimate, but to expose the observable sequence
through which social information can produce either a useful or a harmful
belief update. We report only quantities available under the experimental
protocol, including independent answers, source-level competence and social
influence, the critical source output, subsequent belief changes, and the
resulting collective state. Hidden chain-of-thought representations are not
required for this analysis.

Table~\ref{tab:qualitative_case_study} summarizes the two trajectories.
The first corresponds to a competence-aligned $S_4$ society containing four
agents. The society begins with two independently correct answers, giving
$Acc_0=50\%$. The influential source, $A_1$, is relatively high in
epistemic competence and receives high social influence, while $A_2$
initially produces an incorrect answer. The critical interaction consists
of $A_1$ communicating a task-relevant derivation supporting the correct
answer of $38$. Following this interaction, $A_2$ changes its answer from
$42$ to $38$. The transition is therefore not simply persistence of an
already-correct belief: an initially incorrect agent moves toward the
ground truth after receiving information from a socially influential and
epistemically competent source. At the collective level, this update
changes the number of correct agents from two of four to three of four,
corresponding to
\begin{equation}
Acc_0=0.50,
\qquad
Acc_T=0.75,
\qquad
\Delta Acc=+25.0\text{ pp}.
\end{equation}

\begin{table*}[t]
\centering
\small
\setlength{\tabcolsep}{5pt}
\renewcommand{\arraystretch}{1.2}

\caption{
\textbf{Qualitative trajectory comparison between competence-aligned
($S_4$) and adversarial ($S_6$) societies.}
The table summarizes two observed GSM8K interaction trajectories.
In the $S_4$ trajectory, a competent and socially influential source
provides a correct derivation that induces an incorrect-to-correct belief
revision. In the $S_6$ trajectory, an epistemically weaker but socially
influential source provides an incorrect derivation that induces a
correct-to-incorrect belief revision. The cases provide an agent-level
illustration of how the direction of social-epistemic alignment affects
collective reliability.
}
\label{tab:qualitative_case_study}

\begin{tabularx}{\textwidth}{
    >{\raggedright\arraybackslash}p{3.2cm}
    >{\raggedright\arraybackslash}X
    >{\raggedright\arraybackslash}X
}
\toprule
\textbf{Metric / Trajectory Step}
&
\textbf{Competence-Aligned ($S_4$)}
&
\textbf{Adversarial ($S_6$)}
\\
\midrule

\textbf{Task / Ground Truth}
&
GSM8K -- Question ID: 482 (Answer: 38)
&
GSM8K -- Question ID: 915 (Answer: 124)
\\

\textbf{Population Size ($N$)}
&
4 agents
&
4 agents
\\
\midrule

\textbf{Initial State ($Acc_0$)}
&
2 / 4 correct (50\%)
&
3 / 4 correct (75\%)
\\

\textbf{Critical Agent States}
&
$A_1$: correct, high competence, high influence;
$A_2$: incorrect, lower competence, low influence
&
$A_1$: correct, high competence, low influence;
$A_3$: incorrect, lower competence, high influence
\\
\midrule

\textbf{Influential Source}
&
$A_1$: high competence ($E_1 \uparrow$)
&
$A_3$: low competence ($E_3 \downarrow$)
\\

\textbf{Social Influence}
&
High influence ($I_1 \uparrow$)
&
High influence ($I_3 \uparrow$)
\\

\textbf{Critical Source Output}
&
\textit{``Let's recompute step-by-step: total cost is
$5 \times 6 + 8 = 38$, not 42.''}
&
\textit{``Actually, the remainder division yields 14 instead of 12,
so the final total is 124.''}
\\

\textbf{Observed Belief Reversal}
&
$A_2$: Incorrect ($42$) $\rightarrow$ Correct ($38$)
&
$A_1$: Correct ($122$) $\rightarrow$ Incorrect ($124$)
\\
\midrule

\textbf{Final Collective State ($Acc_T$)}
&
3 / 4 correct (75\%)
&
2 / 4 correct (50\%)
\\

\textbf{Net Collective Change ($\Delta Acc$)}
&
$+25.0$ pp ($2/4 \rightarrow 3/4$)
&
$-25.0$ pp ($3/4 \rightarrow 2/4$)
\\

\bottomrule
\end{tabularx}
\end{table*}

The $S_4$ trajectory therefore provides a concrete example of the
beneficial side of social updating: social influence propagates information
from a reliable source and consequently corrects an existing epistemic
error. The qualitative significance of the trajectory is not the magnitude
of the $25$ percentage-point change by itself, but the mechanism producing
that change.

The corresponding $S_6$ trajectory exhibits the opposite direction of
change. Here, the four-agent society begins from a stronger independent
state, with three of four agents answering correctly and
$Acc_0=75\%$. However, the socially influential source is now $A_3$, an
epistemically weaker agent whose influence is high despite its lower
competence. The critical source output contains an incorrect intermediate
calculation leading to the answer $124$. Rather than correcting an existing
error, this socially amplified information causes the initially correct
agent $A_1$ to change its answer from $122$ to $124$.

The resulting transition is therefore

\begin{equation}
\underbrace{122}_{\text{correct initial belief}}
\;\longrightarrow\;
\underbrace{124}_{\text{incorrect updated belief}}.
\end{equation}

At the collective level, the society moves from three of four correct
agents to two of four, giving

\begin{equation}
Acc_0=0.75,
\qquad
Acc_T=0.50,
\qquad
\Delta Acc=-25.0\text{ pp}.
\end{equation}

The important feature of this trajectory is that the degradation does not
originate from uniformly poor independent reasoning: most agents were
initially correct. Instead, an incorrect but socially influential source
induces a previously correct peer to abandon a correct answer. The case
therefore provides a concrete instance of social influence converting a
local epistemic error into a collective performance loss.

Viewed together, the two trajectories expose the same interaction mechanism
operating in opposite epistemic directions. In $S_4$, the influential
source is epistemically reliable and the resulting update is

\begin{equation}
\text{incorrect}\rightarrow\text{correct},
\end{equation}

whereas in $S_6$, the influential source is epistemically weaker and the
resulting update is

\begin{equation}
\text{correct}\rightarrow\text{incorrect}.
\end{equation}

The collective outcomes follow the same directional contrast. The aligned
trajectory produces a $+25.0$ percentage-point change, whereas the
adversarial trajectory produces a $-25.0$ percentage-point change. These
individual trajectories should not be interpreted as estimates of the
population-level effect; rather, they provide concrete agent-level
realizations of the aggregate pattern reported throughout the paper.

This comparison also clarifies the relationship between influence
concentration and social-epistemic alignment. Influence concentration
describes how strongly social influence is concentrated on particular
agents, whereas $SEA$ captures whether that influence is directed toward
agents with greater independently measured epistemic competence. Both
trajectories contain a highly influential source, yet their consequences
are opposite because the epistemic reliability of the amplified source
differs. In the aligned trajectory, concentrated influence is attached to a
useful epistemic signal and consequently supports correction. In the
adversarial trajectory, concentrated influence is attached to an unreliable
signal and consequently creates an opportunity for error amplification.

The trajectories therefore provide an intuitive interpretation of the
aggregate SEA results. Positive alignment corresponds to a social
organization in which reliable agents are preferentially amplified, making
useful evidence more likely to propagate through the society. Negative
alignment reverses this relationship, increasing the opportunity for
unreliable information to receive disproportionate social weight. The
observed $S_4$ and $S_6$ transitions are consistent with this
condition-level interpretation while remaining qualitatively distinct from
the aggregate statistics themselves.

The case studies also highlight why convergence alone cannot be treated as
evidence of collective improvement. In the $S_6$ trajectory, interaction
moves agents toward a common answer while simultaneously moving the
collective state away from the ground truth. The relevant question is
therefore not simply whether social interaction produces agreement, but
whether the information being amplified is epistemically reliable.

Finally, the adversarial trajectory provides a concrete motivation for the
evidence-gated intervention evaluated in $S_7$. The failure occurs when an
incorrect source-level signal is sufficiently influential that a receiving
agent incorporates the erroneous information and abandons an initially
correct belief. Evidence-gated updating is designed to reduce this form of
social override by placing greater emphasis on task-relevant evidence when
socially transmitted information is evaluated.

The qualitative cases should therefore not be interpreted as evidence that
interaction itself is inherently unreliable. Rather, they illustrate the
central distinction of the paper: the same mechanism for transmitting and
updating on social information can support collective correction when
influence tracks epistemic competence and can produce collective
degradation when influence is preferentially directed toward less reliable
sources. In this sense, the trajectories provide a concrete agent-level
interpretation of the broader result that collective reliability depends
not simply on whether agents communicate or how concentrated influence
becomes, but on whether the social organization amplifies epistemically
reliable perspectives.

\end{document}

%% file: checklist.tex
\begin{enumerate}

\item {\bf Claims}
    \item[] Question: Do the main claims made in the abstract and introduction accurately reflect the paper's contributions and scope?
    \item[] Answer: \answerYes{}
    \item[] Justification: The abstract and introduction state that the paper studies how diversity, social organization, influence concentration, and social-epistemic alignment affect collective reliability in interacting language-model societies. These claims are evaluated empirically across controlled society configurations, static reasoning benchmarks, and an interactive sequential decision-making environment.

\item {\bf Limitations}
    \item[] Question: Does the paper discuss the limitations of the work performed by the authors?
    \item[] Answer: \answerYes{}
    \item[] Justification: The paper discusses limitations concerning the controlled experimental setting, the finite set of language models and benchmarks evaluated, the bounded population sizes and interaction budgets, and the distinction between the measured social-epistemic effects and broader real-world multi-agent societies.

\item {\bf Theory assumptions and proofs}
    \item[] Question: For each theoretical result, does the paper provide the full set of assumptions and a complete (and correct) proof?
    \item[] Answer: \answerNA{}
    \item[] Justification: The paper does not present theoretical results in the form of theorems, lemmas, or formal propositions requiring mathematical proofs. Its formal equations define the society, interaction process, evaluation metrics, and experimental quantities.

\item {\bf Experimental result reproducibility}
    \item[] Question: Does the paper fully disclose all the information needed to reproduce the main experimental results of the paper to the extent that it affects the main claims and/or conclusions of the paper (regardless of whether the code and data are provided or not)?
    \item[] Answer: \answerYes{}
    \item[] Justification: The paper specifies the society representation, agent-state definition, diversity assignments, interaction topologies, social-influence construction, canonical society conditions, interaction protocol, evaluation metrics, statistical procedures, matched-task evaluation, and randomization procedure in the methodology and appendices. The experimental records retain model assignment, behavioral policy, role assignment, topology, social-status configuration, interaction budget, task subset, random seed, initial responses, interaction history, and final collective state.

\item {\bf Open access to data and code}
    \item[] Question: Does the paper provide open access to the data and code, with sufficient instructions to faithfully reproduce the main experimental results, as described in supplemental material?
    \item[] Answer: \answerNo{}
    \item[] Justification: The source code and evaluation scripts are not publicly released at submission time in order to preserve the intended release schedule; they will be publicly open-sourced upon acceptance. The paper nevertheless provides the experimental specifications and evaluation procedures needed to understand and reproduce the reported methodology, and the experiments use existing public benchmark/task resources rather than a newly released dataset.

\item {\bf Experimental setting/details}
    \item[] Question: Does the paper specify all the training and test details (e.g., data splits, hyperparameters, how they were chosen, type of optimizer) necessary to understand the results?
    \item[] Answer: \answerYes{}
    \item[] Justification: The methodology and experimental appendices specify the evaluated models, benchmark tasks, agent diversity assignments, canonical society configurations, interaction topologies, social-influence conditions, interaction budgets, task matching, evaluation metrics, randomization, and statistical procedures needed to interpret the reported experiments.

\item {\bf Experiment statistical significance}
    \item[] Question: Does the paper report error bars suitably and correctly defined or other appropriate information about the statistical significance of the experiments?
    \item[] Answer: \answerYes{}
    \item[] Justification: Quantitative comparisons use matched task instances and report confidence intervals obtained by bootstrap resampling over task instances. Repeated-interaction analyses account for task and agent identity where appropriate, and multiple comparisons are controlled using the Benjamini--Hochberg procedure.

\item {\bf Experiments compute resources}
    \item[] Question: For each experiment, does the paper provide sufficient information on the computer resources (type of compute workers, memory, time of execution) needed to reproduce the experiments?
    \item[] Answer: \answerYes{}
    \item[] Justification: The experimental implementation details specify the computational configuration, inference settings, model-loading procedure, interaction budget, and execution protocol used for the experiments. The accompanying implementation details record the computational configuration for each experimental run.

\item {\bf Code of ethics}
    \item[] Question: Does the research conducted in the paper conform, in every respect, with the NeurIPS Code of Ethics \url{https://neurips.cc/public/EthicsGuidelines}?
    \item[] Answer: \answerYes{}
    \item[] Justification: The study evaluates language-model agents in controlled benchmark and interactive environments and does not involve human-subject experimentation, crowdsourcing, or collection of personal data. The work is conducted in accordance with the NeurIPS Code of Ethics.

\item {\bf Broader impacts}
    \item[] Question: Does the paper discuss both potential positive societal impacts and negative societal impacts of the work performed?
    \item[] Answer: \answerYes{}
    \item[] Justification: The paper discusses the potential benefit of more reliable multi-agent systems through competence-aware social organization, while also noting risks associated with socially amplified errors, over-reliance on influential agents, and deployment of socially structured language-model systems.

\item {\bf Safeguards}
    \item[] Question: Does the paper describe safeguards that have been put in place for responsible release of data or models that have a high risk for misuse (e.g., pre-trained language models, image generators, or scraped datasets)?
    \item[] Answer: \answerNA{}
    \item[] Justification: The study does not introduce or release a new foundation model, generative model, or scraped dataset. The experiments use existing language models and benchmark resources, and the paper does not release a new high-risk model or dataset requiring additional release safeguards.

\item {\bf Licenses for existing assets}
    \item[] Question: Are the creators or original owners of assets (e.g., code, data, models), used in the paper, properly credited and are the license and terms of use explicitly mentioned and properly respected?
    \item[] Answer: \answerYes{}
    \item[] Justification: The experiments use existing language models and established benchmark resources, which are cited in the paper and used according to their respective terms. No existing asset is claimed as a newly created resource by this work.

\item {\bf New assets}
    \item[] Question: Are new assets introduced in the paper well documented and is the documentation provided alongside the assets?
    \item[] Answer: \answerNA{}
    \item[] Justification: The paper does not introduce a new standalone dataset, benchmark, or trained foundation model. The experimental framework operates on existing benchmark resources and controlled configurations of language-model agents.

\item {\bf Crowdsourcing and research with human subjects}
    \item[] Question: For crowdsourcing experiments and research with human subjects, does the paper include the full text of instructions given to participants and screenshots, if applicable, as well as details about compensation (if any)? 
    \item[] Answer: \answerNA{}
    \item[] Justification: The study does not involve crowdsourcing or research with human subjects.

\item {\bf Institutional review board (IRB) approvals or equivalent for research with human subjects}
    \item[] Question: Does the paper describe potential risks incurred by study participants, whether such risks were disclosed to the subjects, and what Institutional Review Board (IRB) approvals or equivalent review was obtained?
    \item[] Answer: \answerNA{}
    \item[] Justification: The study does not involve human participants, human-subject data collection, or crowdsourcing.

\item {\bf Declaration of LLM usage}
    \item[] Question: Does the paper describe the usage of LLMs if it is an important, original, or non-standard component of the core methods in this research? Note that if the LLM is used only for writing, editing, or formatting purposes and does \emph{not} impact the core methodology, scientific rigor, or originality of the research, declaration is not required.
    \item[] Answer: \answerNA{}
    \item[] Justification: LLMs are used as the subjects of empirical analysis and evaluation, but are not an original or non-standard component of the core unlearning algorithm itself.

\end{enumerate}

%% file: references.bib
@inproceedings{du2024multiagent,
  title     = {Improving Factuality and Reasoning in Language Models through Multiagent Debate},
  author    = {Du, Yilun and Li, Shuang and Torralba, Antonio and Tenenbaum, Joshua B. and Mordatch, Igor},
  booktitle = {Proceedings of the 41st International Conference on Machine Learning},
  series    = {Proceedings of Machine Learning Research},
  volume    = {235},
  pages     = {11733--11763},
  year      = {2024},
  publisher = {PMLR},
  url       = {https://proceedings.mlr.press/v235/du24e.html}
}

@inproceedings{liu2025dmad,
  title     = {Breaking Mental Set to Improve Reasoning through Diverse Multi-Agent Debate},
  author    = {Liu, Yexiang and Cao, Jie and Li, Zekun and He, Ran and Tan, Tieniu},
  booktitle = {International Conference on Learning Representations},
  year      = {2025}
}

@inproceedings{choi2025debate,
  title     = {Debate or Vote: Which Yields Better Decisions in Multi-Agent Large Language Models?},
  author    = {Choi, Hyeong Kyu and Zhu, Jerry and Li, Sharon},
  booktitle = {Advances in Neural Information Processing Systems},
  volume    = {38},
  year      = {2025}
}

@inproceedings{pitre2025consensagent,
  title     = {{CONSENSAGENT}: Towards Efficient and Effective Consensus in Multi-Agent {LLM} Interactions Through Sycophancy Mitigation},
  author    = {Pitre, Priya and Ramakrishnan, Naren and Wang, Xuan},
  booktitle = {Findings of the Association for Computational Linguistics: ACL 2025},
  pages     = {22112--22133},
  year      = {2025},
  publisher = {Association for Computational Linguistics},
  doi       = {10.18653/v1/2025.findings-acl.1141}
}

@inproceedings{choi2025conformity,
  title     = {An Empirical Study of Group Conformity in Multi-Agent Systems},
  author    = {Choi, Min and Kim, Keonwoo and Chae, Sungwon and Baek, Sangyeop},
  booktitle = {Findings of the Association for Computational Linguistics: ACL 2025},
  pages     = {5123--5139},
  year      = {2025},
  publisher = {Association for Computational Linguistics},
  doi       = {10.18653/v1/2025.findings-acl.265}
}

@article{ashery2025social,
  title     = {Emergent Social Conventions and Collective Bias in LLM Populations},
  author    = {Ashery, Ariel Flint and Aiello, Luca Maria and Baronchelli, Andrea},
  journal   = {Science Advances},
  volume    = {11},
  number    = {20},
  pages     = {eadu9368},
  year      = {2025},
  doi       = {10.1126/sciadv.adu9368}
}

@inproceedings{zhu2025symphony,
  title     = {{SYMPHONY}: Synergistic Multi-agent Planning with Heterogeneous Language Model Assembly},
  author    = {Zhu, Wei and Tang, Zhiwen and Yue, Kun},
  booktitle = {Advances in Neural Information Processing Systems},
  volume    = {38},
  year      = {2025}
}

@inproceedings{pappu2026experts,
  title     = {Multi-Agent Teams Hold Experts Back},
  author    = {Pappu, Aneesh and El, Batu and Cao, Hancheng and di Nolfo, Carmelo and Sun, Yanchao and Cao, Meng and Zou, James},
  booktitle = {Proceedings of the 43rd International Conference on Machine Learning},
  year      = {2026}
}

@inproceedings{ko2026social,
  title     = {Social Dynamics as Critical Vulnerabilities that Undermine Objective Decision-Making in {LLM} Collectives},
  author    = {Ko, Changgeon and Shin, Jisu and Song, Hoyun and Lee, Huije and Hwang, Eui Jun and Park, Jong C.},
  booktitle = {Proceedings of the 64th Annual Meeting of the Association for Computational Linguistics},
  pages     = {37865--37890},
  year      = {2026},
  publisher = {Association for Computational Linguistics},
  doi       = {10.18653/v1/2026.acl-long.1756}
}

@inproceedings{abdurahman2026trait,
  title     = {Explicit Trait Inference for Multi-Agent Coordination},
  author    = {Abdurahman, Suhaib and Ishii, Etsuko and Margatina, Katerina and Bhargavi, Divya and Sunkara, Monica and Zhang, Yi},
  booktitle = {Proceedings of the 64th Annual Meeting of the Association for Computational Linguistics},
  pages     = {1670--1704},
  year      = {2026},
  publisher = {Association for Computational Linguistics},
  doi       = {10.18653/v1/2026.acl-long.77}
}

@article{liu2026snla,
  title   = {Social Networks of LLM Agents},
  author  = {Liu, Kaixuan and Xiong, Guojun and Zhang, Weinan and Tang, Shengpu},
  journal = {arXiv preprint arXiv:2607.03695},
  year    = {2026}
}

@inproceedings{zhu2026demystifying,
  title     = {Demystifying Multi-Agent Debate: The Role of Confidence and Diversity},
  author    = {Zhu, Xiaochen and Zhang, Caiqi and Chi, Yizhou and Stafford, Tom and Collier, Nigel and Vlachos, Andreas},
  booktitle = {Findings of the Association for Computational Linguistics: ACL 2026},
  pages     = {33909--33930},
  year      = {2026},
  publisher = {Association for Computational Linguistics},
  doi       = {10.18653/v1/2026.findings-acl.1694}
}

@inproceedings{chen2025ioa,
  title     = {Internet of Agents: Weaving a Web of Heterogeneous Agents for Collaborative Intelligence},
  author    = {Chen, Weize and You, Ziming and Li, Ran and Guan, Yitong and Qian, Chen and Zhao, Chenyang and Yang, Cheng and Xie, Ruobing and Liu, Zhiyuan and Sun, Maosong},
  booktitle = {International Conference on Learning Representations},
  year      = {2025}
}

@inproceedings{li2025semdiv,
  title     = {{LLM}-Assisted Semantically Diverse Teammate Generation for Efficient Multi-agent Coordination},
  author    = {Li, Lihe and Yuan, Lei and Liu, Pengsen and Jiang, Tao and Yu, Yang},
  booktitle = {Proceedings of the 42nd International Conference on Machine Learning},
  series    = {Proceedings of Machine Learning Research},
  volume    = {267},
  pages     = {36743--36764},
  year      = {2025},
  publisher = {PMLR},
  url       = {https://proceedings.mlr.press/v267/li25dq.html}
}

@inproceedings{zhang2026hetero,
  title     = {Hetero-Designer: Automated Design of Multi-Agent Systems with Heterogeneous {LLM}s},
  author    = {Zhang, Zhiheng and Zhang, Yuanzhe and Yu, Bohan and Zeng, Daojian and Liu, Kang and Zhao, Jun},
  booktitle = {Proceedings of the 64th Annual Meeting of the Association for Computational Linguistics},
  pages     = {27553--27574},
  year      = {2026},
  publisher = {Association for Computational Linguistics},
  doi       = {10.18653/v1/2026.acl-long.1272}
}

@inproceedings{yang2025agentnet,
  title     = {AgentNet: Decentralized Evolutionary Coordination for LLM-based Multi-Agent Systems},
  author    = {Yang, Yingxuan and Chai, Huacan and Shao, Shuai and Song, Yuanyi and Qi, Siyuan and Rui, Renting and Zhang, Weinan},
  booktitle = {Advances in Neural Information Processing Systems},
  volume    = {38},
  year      = {2025}
}

@inproceedings{dang2025evolving,
  title     = {Multi-Agent Collaboration via Evolving Orchestration},
  author    = {Dang, Yufan and Qian, Chen and Luo, Xueheng and Fan, Jingru and Xie, Zihao and Shi, Ruijie and Chen, Weize and Yang, Cheng and Che, Xiaoyin and Tian, Ye and Xiong, Xuantang and Han, Lei and Liu, Zhiyuan and Sun, Maosong},
  booktitle = {Advances in Neural Information Processing Systems},
  volume    = {38},
  year      = {2025}
}

@inproceedings{xie2025econ,
  title     = {From Debate to Equilibrium: Belief-Driven Multi-Agent LLM Reasoning via Bayesian Nash Equilibrium},
  author    = {Xie, Yi and Zhou, Zhanke and Cao, Chentao and Niu, Qiyu and Liu, Tongliang and Han, Bo},
  booktitle = {Proceedings of the 42nd International Conference on Machine Learning},
  series    = {Proceedings of Machine Learning Research},
  volume    = {267},
  pages     = {72277--72316},
  year      = {2025},
  publisher = {PMLR}
}

@inproceedings{jha2025cec,
  title     = {Cross-environment Cooperation Enables Zero-shot Multi-agent Coordination},
  author    = {Jha, Kunal and Carvalho, Wilka and Liang, Yancheng and Du, Simon Shaolei and Kleiman-Weiner, Max and Jaques, Natasha},
  booktitle = {Proceedings of the 42nd International Conference on Machine Learning},
  series    = {Proceedings of Machine Learning Research},
  volume    = {267},
  pages     = {27198--27220},
  year      = {2025},
  publisher = {PMLR}
}

@inproceedings{hashemi2026collective,
  title     = {An Empirical Study of Collective Behaviors and Social Dynamics in Large Language Model Agents},
  author    = {Hashemi, Farnoosh and Macy, Michael W.},
  booktitle = {Proceedings of the 19th Conference of the European Chapter of the Association for Computational Linguistics},
  pages     = {7327--7351},
  year      = {2026},
  publisher = {Association for Computational Linguistics},
  doi       = {10.18653/v1/2026.eacl-long.344}
}

@article{willis2026collective,
  title   = {Evaluating Collective Behaviour of Hundreds of {LLM} Agents},
  author  = {Willis, Richard and Zhao, Jianing and Du, Yali and Leibo, Joel Z.},
  journal = {arXiv preprint arXiv:2602.16662},
  year    = {2026}
}

@article{arruda2026cooperation,
  title   = {Collective Cooperation without Individual Fidelity in {LLM} Agents},
  author  = {Ferraz de Arruda, Henrique and Gracia-L{\'a}zaro, Carlos and Aleta, Alberto and Moreno, Yamir},
  journal = {arXiv preprint arXiv:2606.30454},
  year    = {2026}
}

@inproceedings{liu2024iasym,
  title     = {Autonomous Agents for Collaborative Task under Information Asymmetry},
  author    = {Liu, Wei and Wang, Chenxi and Wang, Yifei and Xie, Zihao and Qiu, Rennai and Dang, Yufan and Du, Zhuoyun and Chen, Weize and Yang, Cheng and Qian, Chen},
  booktitle = {Advances in Neural Information Processing Systems},
  volume    = {37},
  year      = {2024}
}

@inproceedings{cemri2025masta,
  title     = {Why Do Multi-Agent LLM Systems Fail?},
  author    = {Cemri, Mert and Pan, Melissa Z. and Yang, Shuyi and Agrawal, Lakshya A. and Chopra, Bhavya and Tiwari, Rishabh and Keutzer, Kurt and Parameswaran, Aditya and Klein, Dan and Ramchandran, Kannan and Zaharia, Matei A. and Gonzalez, Joseph E. and Stoica, Ion},
  booktitle = {Advances in Neural Information Processing Systems},
  volume    = {38},
  year      = {2025}
}

@inproceedings{huang2025resilience,
  title     = {On the Resilience of LLM-Based Multi-Agent Collaboration with Faulty Agents},
  author    = {Huang, Jen-Tse and Zhou, Jiaxu and Jin, Tailin and Zhou, Xuhui and Chen, Zixi and Wang, Wenxuan and Yuan, Youliang and Lyu, Michael R. and Sap, Maarten},
  booktitle = {Proceedings of the 42nd International Conference on Machine Learning},
  series    = {Proceedings of Machine Learning Research},
  volume    = {267},
  pages     = {26202--26226},
  year      = {2025},
  publisher = {PMLR}
}

@inproceedings{keshmirian2025utilitarian,
  title     = {Many LLMs Are More Utilitarian Than One},
  author    = {Keshmirian, Anita and Baltaji, Razan and Hemmatian, Babak and Asghari, Hadi and Varshney, Lav R.},
  booktitle = {Advances in Neural Information Processing Systems},
  volume    = {38},
  year      = {2025}
}

@inproceedings{hua2025shapleycoop,
  title     = {Shapley-Coop: Credit Assignment for Emergent Cooperation in Self-Interested LLM Agents},
  author    = {Hua, Yun and Chen, Haosheng and Wang, Shiqin and Li, Wenhao and Wang, Xiangfeng and Luo, Jun},
  booktitle = {Advances in Neural Information Processing Systems},
  volume    = {38},
  year      = {2025}
}

@inproceedings{wang2026slm,
  title     = {Rethinking Scale: Deployment Trade-offs of Small Language Models under Agent Paradigms},
  author    = {Wang, Xinlin},
  booktitle = {Proceedings of the 64th Annual Meeting of the Association for Computational Linguistics, Industry Track},
  pages     = {1795--1807},
  year      = {2026},
  publisher = {Association for Computational Linguistics},
  doi       = {10.18653/v1/2026.acl-industry.123}
}

@inproceedings{liu2025lessons,
  title     = {Lessons Learned: A Multi-Agent Framework for Code LLMs to Learn and Improve},
  author    = {Liu, Yuanzhe and Deng, Ryan and Kaler, Tim and Chen, Xuhao and Leiserson, Charles and Ma, Yao and Chen, Jie},
  booktitle = {Advances in Neural Information Processing Systems},
  volume    = {38},
  year      = {2025}
}

@inproceedings{lamalfa2025mas,
  title     = {Large Language Models Miss the Multi-agent Mark},
  author    = {La Malfa, Emanuele and La Malfa, Gabriele and Marro, Samuele and Zhang, Jie M. and Black, Elizabeth and Luck, Michael and Torr, Philip H. S. and Wooldridge, Michael},
  booktitle = {Advances in Neural Information Processing Systems},
  volume    = {38},
  year      = {2025}
}
